\pdfoutput=1 
\documentclass[debug, preprint, twocolumn]{rmaa}
\usepackage{natbib}

\usepackage{pdfpages}
\usepackage[utf8]{inputenc}

\usepackage{fixltx2e}

\usepackage{color}
\definecolor{URLCOLOR}{rgb}{0.1,0.1,0.7} 
\definecolor{CHANGECOLOR}{rgb}{0.0,0.0,0.0} 

\newcommand\NEW[1]{\textcolor{CHANGECOLOR}{#1}}

\usepackage{aas_macros}
\newcommand{\Cloudy}{\textsc{Cloudy}}
\newcommand{\Hazy}{\textsc{Hazy}}

\newcommand{\cdFilename}[1]{\texttt{#1}}

\font\manual=manfnt at 7pt \def\dbend{\hbox{\raise0.9ex\hbox{\manual\char127\hspace{0.6em}}}}

\providecommand{\e}[1]{\ensuremath{\times 10^{#1}}}

\newcommand\Ion[2]{\ensuremath{\mathrm{#1\,\scriptstyle #2}}}

\newcounter{INTERNALionstage}
\providecommand{\ion}[2]{
  \setcounter{INTERNALionstage}{#2}%
  \Ion{#1}{\Roman{INTERNALionstage}}}

\newcommand{\Av}{A$_{\rm V}$}

\def\gtsim{\mathrel{\hbox{\rlap{\hbox{\lower4pt\hbox{$\sim$}}}\hbox{$>$}}}}
\def\lesssim{\mathrel{\hbox{\rlap{\hbox{\lower4pt\hbox{$\sim$}}}\hbox{$<$}}}}

\def\A{{\rm\thinspace \AA}}
\def\cm{{\rm\thinspace cm}}

\def\erg{{\rm\thinspace erg}}
\def\eV{{\rm\thinspace eV}}
\def\g{{\rm\thinspace g}}

\def\K{{\rm\thinspace K}}
\def\keV{{\rm\thinspace keV}}

\def\micron{\hbox{$\mu$m}}

\def\ps{{\rm\thinspace s^{-1}}}
\def\pcc{{\rm\thinspace cm^{-3}}}

\def\s{{\rm\thinspace s}}
\def\sr{{\rm\thinspace sr}}

\def\ergpscmps{\mbox{$\erg\cm^{-2}\s^{-1}\,$}}

\def\pscm{\mbox{$\cm^{-2}\,$}}

\def\pscm{\mbox{$\cm^{-2}\,$}}
\def\pscmps{\mbox{$\cm^{-2}\, \s^{-1}\,$}}

\def\pressure{\mbox{$\g\, \cm^{-1}\, \s^{-2}\,$}}

\def\la{\mbox{{\rm L}$\alpha$}}

\def\ha{\mbox{{\rm H}$\alpha$}}

\def\hi{\mbox{{\rm H~{\sc i}}}}
\def\hii{\mbox{{\rm H~{\sc ii}}}}

\def\cii{\mbox{{\rm C~{\sc ii}}}}

\def\nii{\mbox{{\rm N~{\sc ii}}}}
\def\oi{\mbox{{\rm O~{\sc i}}}}

\def\oiii{\mbox{{\rm O~{\sc iii}}}}

\def\neiii{\mbox{{\rm Ne~{\sc iii}}}}

\def\siii{\mbox{{\rm S~{\sc iii}}}}

\def\water{\mbox{{\rm H}$_2${\rm O}}}
\def\CO{\mbox{{\rm CO}}}
\def\OH{\mbox{{\rm OH}}}
\def\htwo{\mbox{{\rm H}$_2$}}
\def\hone{\mbox{{\rm H}$^0$}}
\def\hplus{\mbox{{\rm H}$^+$}}

\def\hO{\mbox{{\rm H}$^0$}}

\def\heo{\mbox{{\rm He}$^0$}}
\def\heplus{\mbox{{\rm He}$^+$}}
\def\hePP{\mbox{{\rm He}$^{2+}$}}
\def\cO{\mbox{{\rm C}$^0$}}
\def\cplus{\mbox{{\rm C}$^+$}}
\DeclareMathAlphabet{\vib}{OML}{cmm}{m}{it}

\usepackage{hyperref}
\hypersetup{
    bookmarks=true,         
    unicode=false,          
    pdftoolbar=true,        
    pdfmenubar=true,        
    pdffitwindow=true,      
    pdftitle={The 2013 Release of Cloudy},    
    pdfauthor={G. J. Ferland et al.},     
    pdfsubject={Astrophysics},   
    pdfnewwindow=true,      
    pdfkeywords={keywords}, 
    colorlinks=true,        
    linkcolor=black,         
    citecolor=black,         
    filecolor=black,         
    urlcolor=URLCOLOR           
} 
\newcommand\URL[1]{\href{http://#1}{\nolinkurl{#1}}}

\title{The 2013 Release of Cloudy}
\shorttitle{The 2013 Release of Cloudy}

\author{
G.~J. Ferland\altaffilmark{1}, 
R.~L. Porter\altaffilmark{2}, 
P.~A.~M. van Hoof\altaffilmark{3},
R.~J.~R. Williams\altaffilmark{4},
N.~P. Abel\altaffilmark{5},
\\
M.~L. Lykins\altaffilmark{1}, 
Gargi Shaw\altaffilmark{6},
W.~J. Henney\altaffilmark{7},
P.~C. Stancil\altaffilmark{2}
}
\shortauthor{Ferland et al.}

\altaffiltext{1}{University of Kentucky}

\altaffiltext{2}{University of Georgia}

\altaffiltext{3}{Royal Observatory of Belgium}

\altaffiltext{4}{AWE plc, UK}

\altaffiltext{5}{University of Cincinnati}

\altaffiltext{6}{CEBS, University of Mumbai, India}

\altaffiltext{7}{CRyA-UNAM, Mexico}

\fulladdresses{
\item Nicholas P. Abel: Physics, University of Cincinnati, 4200 Clermont College Drive, Batavia, Ohio, 
45103, USA (npabel2@gmail.com).
\item Gary J. Ferland and Matthew L. Lykins: Physics Department, University of Kentucky, 
Lexington, KY 40506, USA (gary@pa.uky.edu; mllyki3@uky.edu).
\item William J. Henney: Centro de Radioastronom\'\i{}a y
  Astrof\'\i{}sica, Universidad Nacional Aut\'onoma de M\'exico,
  Apartado Postal 3--72, 58090 Morelia, Michoac\'an, M\'exico
  (w.henney@crya.unam.mx).
\item Ryan L. Porter and Phillip Stancil: Department of Physics and Astronomy and Center for Simulational Physics, The University of Georgia, Athens, Georgia 30602-2451, USA (ryanlporter@gmail.com, stancil@physast.uga.edu).
\item Gargi Shaw: Centre for Excellence in Basic Sciences, UM-DAE, University of Mumbai-Kalina Campus, Santa Cruz East, Mumbai-400-098, India (gargishaw@gmail.com).
\item Peter A.~M. van Hoof: Royal Observatory of Belgium, Ringlaan 3, 1180 Brussels,
Belgium (p.vanhoof@oma.be).
\item Robin J.~R. Williams: AWE plc, Aldermaston, Reading RG7 4PR, UK (robin.williams@awe.co.uk)
}

\abstract{%
This is a summary of the 2013 release of the plasma simulation code \Cloudy.  
\Cloudy\ models the ionization, chemical, and thermal state of material that 
may be exposed to an external radiation field or other source of heating, 
and predicts observables such as emission and absorption spectra.  
It works in terms of elementary processes, so is not limited to any particular 
temperature or density regime.  
This paper summarizes advances made since the last major review in 1998.  
Much of the recent development has emphasized dusty molecular environments,
improvements to the 
ionization / chemistry solvers, and how atomic and molecular data are used. 
We present two types of simulations to demonstrate the capability of the code.
We consider a molecular cloud irradiated by an X-ray source such as an Active Nucleus
\NEW{and show how treating EUV recombination lines and the full SED affects
the observed spectrum.}
A second example illustrates the very wide range of particle and radiation density that
can be considered.
}

\resumen{%
  Este es un resumen de la versión 2013 del código \Cloudy{} para simulación de plasmas.
  \Cloudy{} modela el estado térmico, químico y de ionización 
  de la materia que puede ser expuesta a un campo de radiación externa 
  u otras fuentes de calentamiento,
  y predice cantidades observables 
  tales como los espectros de emisión y absorción.
  El código trabaja en términos de procesos elementales y, por lo tanto, 
  no es limitado a un rango particular de densidad o temperatura.
  Este artículo resume los avances logrados desde la última reseña de mayo de 1998.
  Mucho del desarrollo reciente se ha enfatizado en los ambientes moleculares polvosos,
  en mejoras a los resolvedores de ionización/química, 
  y en la forma de uso de los datos atómicos y moleculares.
  Presentamos dos tipos de simulaciones para demostrar las capacidades del código.
  Consideramos una nube molecular irradiada por una fuente de rayos X,
  por ejemplo, un Núcleo Activo,
  e ilustramos el efecto sobre el espectro observado de líneas de recombinación EUV 
  y de la distribución espectral completa de la radiación.
  Un segundo ejemplo destaca el rango tan amplio de densidad de partículas y de radiación que se puede considerar.
}

\addkeyword{atomic processes}
\addkeyword{galaxies: active}
\addkeyword{methods: numerical}
\addkeyword{molecular processes}
\addkeyword{radiation mechanisms}

\reviewarticle
\renewcommand\reviewname{Review Article}

\SetFirstPage{1}
\SetYear{2013}

\begin{document}

\RescaleLengths{1.1}

\maketitle
\clearpage
{\small
  \reviewtoc
  \bigskip
}


\section{Introduction}

Most quantitative information we have about the cosmos comes from spectroscopy. 
In many astronomical environments the density is too low for equilibrium thermodynamics to apply,
so the ionization, molecular state, level populations, kinetic temperature, 
and the resulting spectrum are the result of a host of microphysical processes.  
As a result  the spectrum reveals much about the properties of an object, 
but it also means that modeling this detail is complex.  
Analytical results are only possible in certain limits, so numerical simulations must be used.  
Texts that review this field include 
\citet{Spitzer1978},
\citet{DopitaSutherland03},
\citet{Tielens2005}, 
\citet{AGN3} (hereafter AGN3), 
and \citet{Draine.B11Physics-of-the-Interstellar-and-Intergalactic-Medium}.

\Cloudy\footnote{www.nublado.org}  is an open source plasma simulation code that is designed to 
simulate conditions in a non-equilibrium gas, and predict its spectrum.  
The code incorporates physical processes from first principles, as much as possible.  
The goal is to simulate the ionization, level populations, molecular state, and thermal state, 
over all extremes of density and temperature.  
Our approach, working from fundamental processes, means that \Cloudy\ can be 
applied to such diverse regions as the corona of a star, the intergalactic medium, 
or the accretion disk near the supermassive black hole in a luminous quasar.  
As a result, the code is widely used, with nearly 200 papers citing its documentation each year.  
The diversity of problems it can address is a testimonial to the importance 
of treating the atomic physics at an elementary level.  

Processor power has always limited our ability to simulate detailed microphysics.  
Improved computers and advances in atomic and molecular physics allow a better simulation.  Improved numerical methods or coding techniques make the solutions more robust.  
These advancements to the fidelity of the simulation improve our insight into the 
inner workings of astronomical objects.  
Because of these changes, \Cloudy, like most software, goes through a 
development / release cycle. 
Our goal is to make major updates every two or three years. 

This paper is a progress report on the improvements to \Cloudy\ since the last major review, 
\citet{Ferland1998}, referred to as F98 in the following. 
Although the code's download includes extensive documentation that is continuously updated, 
there has not been a recent major review.  
We rectify that need here.

\Cloudy's development leading up to the 1998 review had emphasized the UV, optical, 
and IR spectra of ionized gas. 
The simulations have been extended to fully molecular regions 
with predictions of the associated IR and radio emission since then.  
The code can handle a broad range of physical state, 
from predominantly molecular to fully ionized, a broad range of densities 
from the low density limit to roughly $10^{15} \pcc$, and temperatures ranging from 
the CMB to $10^{10} \K$.

In the present paper, we summarize the major advances in the code since F98.  
Most of this work has been documented in past papers, which we cite.  
In the interests of brevity we only give references to the relevant papers with a brief
summary of the advances they document. 
We discuss some technical details about the code, its operation, and installation.
We present several calculations that show the range of applicability of \Cloudy.
These include the properties of the ionized, atomic, and molecular
gas produced by the radiation field of an Active Galactic Nucleus (AGN) and a demonstration of
the range of particle and radiation density that can be considered.
We conclude with an outline of future development directions.

\section{The physics of ions, molecules, and grains}

\Cloudy\ was originally designed to simulate the dense gas found near the black hole 
and accretion disk in \NEW{AGN\@.}  
The so-called ``Broad Line Region'' (BLR) mainly emits in the UV and optical, 
and these spectral regimes were the original focus.  
Some investigations are
\citet{Rees1989}, \citet{FerlandPeterson1992}, and   \citet{Baldwin1995}.
Lower density gas, at larger distances from the center, 
produces the ``Narrow Line Region'' (NLR) spectrum, which includes 
a range of forbidden and permitted lines in the UV, optical, and IR.  
The study by \citet{Ferland1983} is an example.  
Such investigations drove the development of \Cloudy, as summarized in the 1998 review, 
although Photo-dissociation Region (``PDRs'') and 
X-ray Dissociation Regions (``XDRs'') were also simulated 
\citep{Ferland1994, FerlandFabian2002}.  

The following sections describe our improvements in the treatment of ions, molecules, and grains 
since F98,
by citing those papers which introduced the advances.  This is not meant to be a
comprehensive review of the literature, rather, only a description of advances to 
\Cloudy\ since the last review.

\subsection{Structure of the H-like and He-like iso-electronic sequences}

Hydrogen and helium are the most common elements in the universe and, as a result, 
need to be treated with the greatest precision.  
Their structure is different from most heavy elements,  with a first excited level 
at (1-1/$n^2$) = 0.75 of the ionization energy and more highly excited levels 
close to the ionization limit.  
By contrast, the heavy elements can have many low-lying levels. 
This means that, for most temperatures, lines from H and He -- like species 
will have a strong recombination component, while lines of many-electron systems 
will be predominantly collisionally excited.
The structure of the H and He -- like iso sequences also means that a significant number
of excited states are needed for the model atoms to correctly go to LTE in the high
particle or photon limits.

Steve Cota developed the original model of \ion{H}{1}, \ion{He}{1}, and 
\ion{He}{2}\ emission in \Cloudy\ 
as part of his PhD thesis \citep{1987PhDT.........7C}.  
He also developed an approximate treatment of three-body recombination for the heavy
elements, described in the next section.
This was extended by Jason Ferguson in his PhD thesis
\citep{Ferguson.J97Spectral-Simulations-and-Abundance-Determinations, FergusonFerland1997} 
to include more levels, 
as processor power increased.
These models, which involved 15 levels with a number of higher pseudo states, 
were significant time sinks on the computers at that time.

Today's unified model of the H and He iso-electronic sequences was developed as part of 
Ryan Porter's thesis.  
Because the high charge states occur in hot gas and the line energies scale as $Z^2$, 
these iso-sequences sort themselves into two spectral regions.  
\ion{H}{1}, \ion{He}{1}, and 
\ion{He}{2}\  produce strong lines in the optical 
while \ion{C}{5}, \ion{C}{6}, \ion{O}{7}, \ion{O}{8}, \ion{Fe}{25}, \ion{Fe}{26}, etc 
produce lines in the X-ray.  
Despite these differences, the physics has many similarities.  
These sequences are treated with a common code base, 
which results in greater simplicity and reliability.

\citet{Porter2005} and  \citet{Bauman2005} describe the model of  \ion{He}{1}\ emission.  
\citet{PorterFerland2007} describe ions of the He sequence, 
which mainly emits in the X-Ray.  
\citet{LuridianaEtAl09} summarize expansions to the H-like sequence while
\citet{PorterFerlandMacAdam2007}, 
\citet{Porter.R09Uncertainties-in-theoretical-HeI-emissivities:-HII-regions},
and
\citet{Porter.R12Improved-He-I-emissivities-in-the-case-B-approximation}
discuss uncertainties and more recent improvements in the atomic data for \ion{He}{1}.

\begin{figure}[t]
\includegraphics[width=\linewidth]{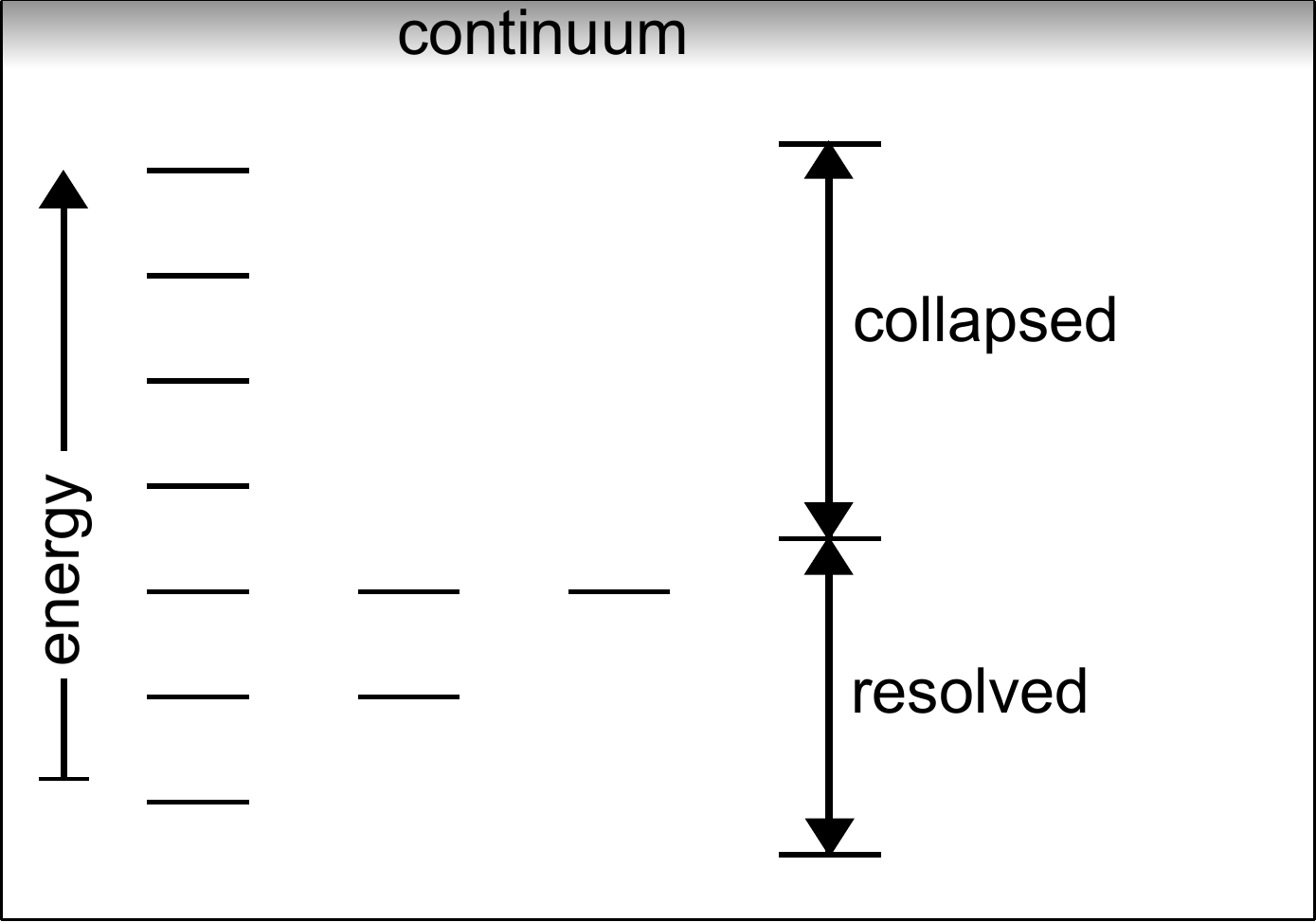}
\caption[Resolved and collapsed levels in the iso-sequence mode atoms]{The iso-sequence atomic level structure consists of 
low-lying $nl$ resolved ($nls$ for He-like) terms
with $l-$mixed ($ls$ for He-like) collapsed configurations.
The number of resolved and collapsed levels can be specified
when the simulation parameters are established.
\label{fig:LevelsResolvedCollapsed}}
\end{figure}

A schematic representation of one of the elements of the H-sequence is shown in 
Figure~\ref{fig:LevelsResolvedCollapsed}.  
The He-sequence has similar structure except that it is resolved into singlets and triplets 
with twice as many levels. 
Principal quantum number increases upward with the continuum at the top
and $nl$ terms are indicated from left to right.  
Lower $n$ configurations are resolved into $nl$ terms, the ``resolved'' levels.  
Above a certain quantum number $l$-changing collisions become fast enough 
to guarantee that $l$ terms are populated according to their statistical weight 
within the $n$ configuration \citep{PengellySeaton1964}.  
Such higher levels are treated as ``collapsed levels'' which are $nl$ mixed.

This treatment includes ions of all elements up to Zinc.  
The models include photoionization / recombination, collisional ionization / 3-body recombination, 
to all levels, and collisional and radiative processes between levels,
so behave correctly in the low density limit and go to LTE at high densities 
or exposed to a true blackbody radiation field \citep{Ferland1988}.  
Line trapping, collisions, continuum lowering, and absorption of photons by continuous opacities, 
are all included as general processes \citep{Rees1989}.

The user can adjust the number of resolved and collapsed levels modeled 
when the simulation is specified.  
The spectrum is predicted with great precision, an accuracy of better than 1\%, 
when a larger number of levels are used 
\citep{Porter.R12Improved-He-I-emissivities-in-the-case-B-approximation}.  
This comes at the cost of increased execution times.  
Smaller models are often used for simulations of clouds with significant column densities 
due to the compute time required.  
The default treatment includes the greatest number of levels for H, 
and increasingly smaller numbers of levels for He, common second row elements like C and O, 
Fe, and the remaining low abundance elements. 

\subsection{Structure of other ions}

\Cloudy\ includes all ions of the lightest thirty elements.
\citet{Lykins.M12Radiative-cooling-in-collisionally-and-photo} summarize recent developments
in the treatment of the ions that are not part of the H and He iso-sequences.

An equivalent two-level system is assumed in modeling the ionization balance of ions
of the Li-like and multi-electron iso sequences.
Photoionization and collisional ionization from the ground configuration is balanced by recombinations
to all levels.
This assumes that nearly all populations are in ground, an approximation which is valid
for moderate densities and the low temperatures usually found in photoionization equilibrium.
Photoionization cross sections are from the Verner database 
\citep{VernerFerlandKorista1996} while
radiative and dielectronic recombination rates are \NEW{largely} from the Badnell web 
site\footnote{\url{http://amdpp.phys.strath.ac.uk/tamoc/DATA/}} as
described in \citet{BadnellEtAl03} and \citet{Badnell06},
supplemented by data calculated as described by \citet{VernerFerland1996}.
Charge exchange ionization and recombination rates are taken from 
an updated version of the \citet{Kingdon1996} database.

The treatment of inner-shell processes, including line emission following removal of an
inner-shell electron, largely follows F98.  
As described below, the treatment of multiple electron ejection, which had
followed \citet{Weisheit1972}, 
is now generalized to non-adjacent stages of ionization \citep{HenneyEtAl05}.  
Ionization and recombination coupling non-adjacent ion states 
can also be important in grain surface recombination. 

This treatment is approximate at high densities and for temperatures which are 
a significant fraction of the ionization potential of a species.  
The use of summed recombination rate coefficients in effect assumes
that all recombinations eventually populate the ground level.  
Ionization processes out of excited levels are neglected.
Both are no longer true if the density is large enough for excited levels to
play an important role. 
Expanding the treatment of these species to the approach now used for the
H and He iso sequences is a high priority for future development.

Bound levels of each ion are treated with a variety of models.
We are enhancing the code to use external databases as much as possible.
We have the ability to read the Chianti atomic 
database\footnote{CHIANTI is a collaborative project involving the NRL (USA), the Universities of Florence (Italy) and Cambridge (UK), and George Mason University (USA).}  
\citep{Dere.K97CHIANTI---an-atomic-database-for-emission, Landi2012}  and
this release of \Cloudy\ includes Version 7.0.
Chianti is used for most ions of Fe while for other species we mainly use our internally developed 
atomic database. 
\citet{Lykins.M12Radiative-cooling-in-collisionally-and-photo} shows that our original
database, which is embedded in the C++ source, is in good agreement with Chianti.

Additionally we are starting to develop our own database, ``Stout''.
We will add new models of ions or molecules to this second database, 
and make it publicly available along with \Cloudy.

\subsection{Molecular chemistry}
\label{molchem}

\Cloudy\ initially included the chemistry network described by \citet{Black1978}
which was expanded to \NEW{treat} PDRs and XDRs as described by \citet{Ferland1994}.
Nick Abel \NEW{carried out} a massive upgrade to the heavy-element chemistry network
as part of his PhD thesis, described in \citet{Abel2004}.  
Later refinements are discussed in \citet{Abel2005},
\citet{Shaw2005}, and \citet{Shaw2006}.  
Appendix~A of \citet{Abel2005} gives details of the numerical methods along with differences 
between UMIST and \Cloudy\ reaction rates.  
\Cloudy\ had predicted column densities for about 20 heavy element molecules, 
consisting of C and O atoms.
It could not calculate physical conditions deep in a 
PDR or a molecular cloud, where most gas phase C, N, and O 
is in the form of molecules, due to numerical instabilities in the chemistry solver then used.   
The upgraded chemistry solver has no restrictions, as described in sections below.
\Cloudy\ now calculates the chemical abundance of 83 molecules using a network including $\sim 10^3$ 
chemical reactions involving molecules containing H, He, C, N, O, Si, S, and Cl atoms.  The network adjusts automatically when elements or species are disabled.
  
Most reaction rates come from the UMIST 2000 database 
\citep{Le-Teuff.Y00The-UMIST-database-for-astrochemistry-1999}
as updated for the Leiden workshop and described by \citet{Roellig2007}.
We also predict the freeze-out of \water, \CO, and \OH\ on grains, 
using the data given in \citet{Hasegawa1993}.  
Both time-steady and time-dependent chemical evolution calculations are possible.  

The effects of cosmic rays and suprathermal secondary electrons can be very important
in molecular regions \citep{1968ApJ...152..971S} and are treated
as described by \citet{FerlandMushotzky1984} and \citet{FerlandFabianEtAl2009}.
Non-thermal particles can both excite and dissociate the gas.
\citet{Ferland1994} and \citet{FerlandFabian2002} describe the treatment
of chemistry in an X-ray dominated filaments in cool-core galaxy clusters.
\citet{Shaw2008} describe an application to the local ISM.

The treatment of \htwo\ is fully self consistent with its surroundings
\citep{Shaw2005}.
Formation on grain surfaces is treated using the derived grain properties and
the \citet{Cazaux.S04H2-Formation-on-Grain-Surfaces} catalysis rates.
Destruction by photoexcitation in electronic states is treated by computing
the radiation field at each point.
This includes line self shielding, the attenuation
by continuous opacity sources such as grains, photoelectric opacity of the heavy elements,
and Rayleigh scattering,
and emission by the cloud itself.
\hi\ \la, the \ion{He}{1} resonance lines, and electronic lines of \htwo\ are especially important sources 
of higher-energy photons.

\subsection{Molecular excitations and spectra}

We next describe our treatment of molecular emission processes.
Gargi Shaw developed our model of the \htwo\ molecule,
the most common molecule in the Universe,
as part of her PhD thesis \citep{Shaw2005}.
All levels within the ground electronic state are included
\citep{Dabrowski.I84The-Lyman-and-Werner-bands-of-H2}, along with
all electronic excited states on the Meudon web 
site\footnote{\url{http://molat.obspm.fr/index.php?page=pages/Molecules/H2/H2can94.php}}
and \citet{Abgrall.H94The-Bprime1Sigmau-rarr-X1Sigmag-and-D1Piu}.
Collision rates are from \citet{Wrathmall2007}.
\citet{Lee.T05Close-coupling-calculations-of-low-energy-inelastic,
Lee.T06State-to-state-rotational-transitions-in-H2H2,
LeeH2H22008},
\citet{Shaw.G09Rotationally-Warm-Molecular-Hydrogen},
and
\citet{Gay.C12Rovibrationally-Resolved-Direct-Photodissociation}
summarize more recent updates.

Grain formation pumping of \htwo\ can be treated using several theories.
\citet{Shaw2005} provides more details.
We use the \citet{Takahashi2001} results by default.

Line emission for molecules heavier than \htwo\ is predicted using the
level energy, collision, and radiative data in the LAMDA database
\citep{Schoier.F05An-atomic-and-molecular-database-for-analysis}.  The
molecular excitation are solved simultaneously and self-consistently
with the global environment.  This includes grain properties,
including emission, so molecular pumping in the infrared continuum is
automatically included (if the pumping lines are present in the molecular data), along with attenuation by grains and other continuous opacity sources.
Line optical depths are computed for each point in the cloud, and the full radiative transfer
performed as described below.

\subsection{Grains}

Grains have been included in \Cloudy\ since the very beginning
\citep{Martin1980}, and the basic physical processes are summarized in
Appendix~C of \citet{BFM}. \citet{VanHoof2004} describe the main
improvements made since \citet{BFM}.

The grain model in \Cloudy\ includes all relevant processes: absorption and
scattering of light including stochastic heating effects, the photoelectric
effect including Auger emissions in X-ray environments, collisional charging
(electrons as well as atomic ions), thermionic emissions, collisional energy
exchange between the grains and the gas, and a calculation of the grain drift
velocity. We also consider molecular freeze-out on grain surfaces and grain
surface reactions. These are discussed in Section~\ref{molchem}.

The local radiation field, including the attenuated incident Spectral Eenergy Distribution (SED) and the spectrum
emitted by the cloud (generally \la\ is the most important) are all treated self-consistently,
with \NEW{gas and} grains providing the opacity affecting the light, and the light affecting the grain properties
following absorption.
The code fully treats stochastic heating of grains using a robust and efficient algorithm (which is a
comprehensively upgraded version of a code originally written by K. Volk), implementing an
improved version of the procedure described in \citet{Guhathakurta1989}. This includes an
approximate treatment for stochastic heating by particle collisions. Combined with
resolved size distributions, this will lead to a much more realistic modeling of the grain emission
under all circumstances. The stochastic heating code needs a grain enthalpy function. Supported functions are
taken from \citet{Guhathakurta1989}, \citet{Dwek1997}, and \citet{Draine2001}.

The grain physics includes the advances discussed by \citet{Weingartner2001a}
and \citet{Weingartner2006}. The latter provides a realistic model for grains in X-ray
environments, including Auger emissions by the grains.
The grain
model has a detailed treatment of the photoelectric effect and collisional processes, and includes
thermionic emissions.
The charge for each grain constituent is determined self consistently with the local radiation
field and gas properties, 
using the hybrid grain charge model described by \citet{VanHoof2001} and \citet{VanHoof2004}.
The grain charge is included in the overall charge balance of the system, which has a
significant effect on the modeling results for molecular regions \citep{Abel2008}.
Charge exchange on grain surfaces is treated following \citet{Draine1987} assuming that
electrons will be exchanged between the grain and the colliding particle until a minimum
energy state is reached. The grain drift velocity w.r.t.\ the gas is calculated using
the theory from \citet{Draine1979}.

Extensive comparisons in collaboration with Joe Weingartner done in 2001 show that the photoelectric heating rates and collisional cooling rates predicted by \Cloudy\ agree very well with
the results from the \citet{Weingartner2001a} model for a wide range of grain sizes (between
5~\AA\ and 0.1~$\mu$m), and using various choices for the incident radiation field. 
A detailed discussion of
this comparison can be found in \citet{VanHoof2001}.

We include an embedded model for Mie scattering \citep{Mie1908}, which uses refractive index data for each
material type.  It is based on the code described in \citet{Hansen1974} and was later
modified by P.G.\ Martin.
New grain materials can be included by specifying the relevant refractive index data
or by supplying opacity tables.
The user specifies the size-distribution function and the number of grain size bins.
A number of size-distribution functions are available as built-in functions, including
the ones described in \citet[hereafter MRN]{Mathis1977}, \citet{BFM}, and \citet{Abel2008}.
\Cloudy\ will then ``compile'' the grain data to create scattering and absorption cross sections,
the scattering asymmetry factor, and the inverse attenuation length for each material type and grain size
as a function of frequency.
A number of pre-compiled grain types are included in the code distribution.

In practice, it is difficult to introduce new grain materials because of the wavelength range
considered by the code.  Laboratory data can extend from the IR into the UV, 
but little data exists shortward of $\sim 0.1 \mu$m.  
Theoretical relationships between the bulk grain properties and the photoionization
cross sections of the constituent atoms can be used to create complete refractive index data 
for grain materials when combined with the Kramers-Kronig relations.
Refractive index data for astronomical silicate, amorphous carbon, and graphite, based on
\citet{Martin1991}, \citet{Rouleau1991}, and \citet{Laor1993} are included in the \Cloudy\ download.
Opacity data for PAHs from Volk (private communication, based on data from
\citealp{Bregman1989}, \citealp{Desert1990}, and \citealp{Schutte1993}),
\citet{Li2001}, and \citet{Draine2007} are built into the code.

The Mie code includes the possibility to mimic mixtures of materials in grains
using effective medium theory (EMT). The following EMT recipes are supported
by \Cloudy: \citet{Bruggeman1935}, \citet{Stognienko1995}, and
\citet{Voshchinnikov1999} (based on the theory in \citealp{Farafonov2000}).
The first two are appropriate for randomly mixed grain materials, while the
latter is intended for layered grains. \Cloudy\ includes a refractive index
file for vacuum which enables modeling fluffy grains when combined with other
materials using an EMT.

\citet{Abel2008} discusses how our treatment, which is based on elementary processes
as \NEW{far} as possible, affects results for PDR simulations.
Conventional PDR codes use precomputed integrals of photoionization or photodissociation 
cross sections over an SED representative of the Galactic starlight background.  
This rate is assumed to be a function of the radiation field scaled to the intensity of the Galactic
background, and the visual extinction \Av.  
In contrast we explicitly integrate stored cross sections over the local radiation field
for those processes which have energy-specific data.
These include photo rates for all atoms, ions, and grains, and some molecules, most notably \htwo.
The incident radiation field is attenuated by the computed gas and dust opacity.
Our treatment can handle any grain opacity distribution or abundance, or SED shape.

The grain scattering theory predicts the scattering asymmetry factor $g$,
which is the average of the cosine of the scattering angle of an incident photon
\citep{Martin1979}.
When a point source like a star is viewed through an intervening cloud at a large distance from the star,
photons scattered by even a small amount are lost from the beam and cannot reach the observer.
So in this case scattering attenuates the radiation field by the full opacity, $\alpha_{scat}$.
We refer to this, the quantity measured by observations of stellar extinction,  as the point-source extinction.
On the other hand, when modeling a cloud irradiated by a star, photons that are scattered
in a forward direction can still propagate into the next radial zone of the model and are
therefore not lost from the radiation field \citep{Martin1979, BFM, AGN3}.
In \Cloudy\ this is approximated using an effective opacity $\alpha_{scat} (1-g)$,
which is said to discount forward scattering. We refer to this as the extended extinction.
Both extended and point-source extinctions are reported by the code,
but the user should be aware that the PDR literature always uses the point-source extinction.

\subsection{Line transfer}

The treatment of line transfer is largely unchanged from F98.
The escape probability formalism is used,
using the framework given by, among others,
\citet{Irons.F78On-the-equality-of-the-mean-escape-probability},
\citet{FerlandElitzur1984},
\citet{Rybicki.G84Escape-probability-methods-pp-21-64},
\citet{Kalkofen.W84Methods-in-radiative-transfer},
\citet{Netzer1985},
\citet{Elitzur.M85Line-fluorescence-in-astrophysics},
\citet{Elitzur1986},
\citet{Kalkofen1987}, 
\citet{Ferland1992},
and
\citet{Elitzur1992}.

All permitted, and many forbidden, lines are transferred using a common approach.  
With this, processes such as line trapping and thermalization, pumping by the
local radiation field, line destruction by background opacities such as
photoelectric or grain absorption, are included for the $\sim 10^5 - 10^6$ 
atomic and molecular lines considered by the code.
Line processes couple into the model atoms, which can go over to
the correct thermodynamic limits when exposed to a blackbody
\citep{Ferland1988}.
The simulation is done self consistently, with feedback between various constituents taken into account.
For instance, thermal grain infrared emission can fluoresce atoms or molecules
within the cloud, and grain opacity impedes the propagation of emission lines out of the cloud.

As part of the maintenance and improvement of the underlying atomic database
we have incorporated the UTA data computed by
\citet{Kisielius2003} and updated by Ferland et al. (2013, ApJ submitted).
The radiative rates are very large and the corresponding radiative damping parameters
can be substantially greater than 1.
To improve the accuracy of predictions for such lines with high
damping parameters, the Voigt function used in line transfer
calculations now uses the fast and accurate implementation of
\citet{Wells1999}, with some modifications to improve performance in
the limit of small damping using the theory presented by
\citet{Hjerting1938}.

The code includes a number of hyperfine structure lines 
\citep{Goddard.W03Hyperfine-Structure-Emission-and-Absorption}.
We have put a major emphasis into the physics of the \hi\ 21 cm line
\citep{Shaw.G05Role-of-molecular-hydrogen-in-star-forming}.
The line is usually optically thick, and pumping by \hi\ \la\ is treated as in
\citet{Deguchi1985}.
\citet{Pellegrini2007} show how feedback between the stellar SED and
fluorescent excitation of \hi\ \la\ alters the 21 cm optical depth and spin temperature
in the M17 \hii\ region.
 
\subsection{Solution of material state}

Cloudy solves for self-consistent \NEW{populations of electrons, ionized
species, populations of excited levels of atoms, ions, and molecules, 
molecular chemistry}, and the charge state and temperature distribution of a spectrum of
grains.  This population balance is then used by outer solvers for the
material temperature, pressure (when required) and radiation transfer,
and used to predict the resulting spectrum.  As described in the
previous sections, the range of physics treated has expanded
significantly since the last review.

At the simplest level, the method of solution for the populations
remains similar to that used previously, with the populations of
different systems solved for in an iterative loop.  However, the
increasing level of coupling between species has required significant
work to be done on improving the robustness of convergence and the
self-consistency of the solutions.  The broad range of physics which
Cloudy solves in a self-consistent fashion means that the resulting
system has a very wide range of physical timescales, a significant
challenge for any numerical scheme; the status of the code as an open
resource for the astrophysical community means that this challenge
must be met by means which require no user intervention.  Details of
this work will be described in detail in Williams et al.\ (2013, ``Hierarchical
physics'', in preparation), so in the present paper we will summarize
the overall approach.

The molecular chemistry is now fully consistent with the ionization
balance.  This is done by scaling chemical reactions including atoms
and positive ions to have rates dependent on a combined species which
is the sum of the ions within the chemical network solver.  The
resultant effects of the chemistry on the ionization balance are
allowed for by having the molecular network calculate the net source
and sink rates for these ions from molecular chemistry, which are then
included in the ionization balance solver.

The nonlinear system for the molecular chemistry is solved using an
adaptive timestep implicit solver: in typical usage, this is set up to
run the chemical balance to full late time equilibrium.  The molecular
chemistry and ionization solvers have been adapted to allow the
solution of time-dependent and steady-state advective flows, using the
approach described by \citet{HenneyEtAl05}.  
\NEW{Source and sink terms are inserted} in the existing equilibrium solvers to
adapt them to find the solution of a backward Euler implicit system
for the time advance of the state.  This allows the code to take
advantage of the existing equilibrium solvers with minimal change,
exploiting these to provide an implicit solution for the time advance,
which is necessary given the wide range of physical timescales which
operate in the systems which are modeled.  \NEW{The steady flow model was applied}
by \citet{HenneyEtAl05} to treat the structure of \hii\ region
photo-evaporation fronts, and extended by \citet{Henney2007} to the
molecular knots within planetary nebulae, including the PDR ahead of
the \hii\ region.

Work has also gone into improving the robustness of handling of
processes which couple different parts of the system sufficiently
strongly that a simple iterative scheme was slowly convergent.  The
particular cases where this has been found to be an issue is in the
handling of the resonant O/H charge transfer process
\citep{Stancil.P99Charge-transfer-in-collisions-of-O-with}, the rate of
which can by far dominate direct ionization processes for either ion,
and the handling of Rydberg levels which are more strongly coupled to
the continuum than the base ion.  A simple solution acceleration
approach has been found to be sufficient to allow rapid convergence,
details of which will be given in Williams et al (in preparation).

In addition, the solvers for the electron density and temperature
have been completely rewritten and are now much more robust and
reliable than the previous versions. These changes are especially
important for modeling extreme environments: very cold regions on
the one hand (PDRs and molecular regions) and very hot on the other
(extreme X-ray environments such as disks surrounding a black hole).

\subsection{Momentum balance and the equation of state}
\label{sec:eos-momentum}
\Cloudy{} allows great flexibility in specifying 
the spatial distribution of the gas (and grain density).
In addition to constant density and various ad~hoc density laws
(e.g., power law),
it is also possible to allow \Cloudy{} 
to solve for a self-consistent density distribution 
based on momentum balance. 
In the simplest case of a static configuration with no external forces
this reduces to the requirement of constant total pressure, 
whereas with the addition of an external force, 
such as continuum radiation pressure \citep{BFM}
or gravity \citep{AscasibarDiaz2009},
it becomes hydrostatic \NEW{or magnetostatic} equilibrium. 
\Cloudy{} also allows the further generalization to the case of
dynamic equilibrium in the presence of 
steady-state gas flow \citep{HenneyEtAl05}. 

In addition to the thermal pressure of the gas, 
\Cloudy{} also considers other contributions to the total pressure. 
These include trapped resonance line radiation \citep{Elitzur1986},
ordered and disordered magnetic fields \citep{HenneyEtAl05, Pellegrini2007},
cosmic rays \citep{Shaw.G09Rotationally-Warm-Molecular-Hydrogen},
and the Reynolds stress due to turbulent motions.

\section{Some computational details}

\subsection{An Open Source Project}
\Cloudy\ is openly available on the web at the site 
\URL{www.nublado.org}.  
This includes the full source,
the atomic and molecular data needed for \Cloudy\ to operate, 
its documentation \Hazy, and an extensive suite of test cases.
The test suite illustrates how to use the code, its range of validity, and includes embedded monitors that
confirm that the code is operating correctly.
All previously released versions of \Cloudy\ are available on the web site, with most
stored in an openly accessible \textit{Subversion} repository.  
The distribution is subject to a BSD style license.

Although this is the first major review since F98,
\Cloudy\ has been continuously developed, as witnessed by the papers cited above.
New versions are released every two to three years, at the conclusion of a period
of development which focused on particular aspects of the simulations. 
This is the seventh major release since F98.
The \URL{www.nublado.org}  site gives the full history.

At the time of F98, the code was $\sim 9\times 10^4$ lines written in Fortran 77, 
with some portable extensions. 
It transitioned to C in 1999 and C++ in 2006.  
Today \Cloudy\ consists of roughly $2.1\times 10^5$ lines of C++ source code.
In recent years, the size of the code has stabilized, 
as work to extend its scope is balanced by the move from the inclusion of 
physics data within the source coding towards the use of external database files.

The code has a broad user community.
As described below, we ask that users reference \NEW{the current} paper if the code is used in a publication.
At the time of this writing there are nearly 200 citations to F98 or the code's documentation each
year.
We maintain a
\href{http://tech.groups.yahoo.com/group/cloudy_simulations}{discussion board}\footnote{
\url{http://tech.groups.yahoo.com/group/cloudy_simulations}} 
where users can ask questions and where we post
announcements of updates to the code.

\subsubsection{Material available on \URL{www.nublado.org}}

\URL{www.nublado.org}, the code's web site, gives complete
access to files and information about \Cloudy.
The site, built using \href{http://trac.edgewall.org/}{trac},
gives top level links to a variety of items.  These include:

\begin{itemize}

  \item \href{http://www.nublado.org/wiki/StepByStep}{Step by step instructions} 
  for downloading, building, and running the code.  
  We provide links to a large tarball for the download, and makefiles are used to build
  the code.  

  \item \href{http://www.nublado.org/wiki/StellarAtmospheres}{Stellar atmospheres}.  
  As discussed in Section \ref{sec:StellarAtmospheres} below,
  it is possible to use grids of stellar atmospheres in deriving the incident radiation field.
  This page gives more details and links to available grids. 
 
  \item \href{http://www.nublado.org/wiki/KnownProblems}{Known problems}, and 
  \href{http://www.nublado.org/wiki/HotFixes}{hot fixes}.  No code is perfect.  
  Users should post questions or bug reports on the 
  \href{http://tech.groups.yahoo.com/group/cloudy_simulations}{discussion board}.
  We provide a list of known problems.
  These are deficiencies which we know about, but which have not been
  fixed in the current version.
  Hot fixes are small changes to the source code which will fix problems discovered since the last version
  of the code was released.  They should be applied to the code source before building it.
  
  \item The \href{http://www.nublado.org/wiki/RevisionHistory}{revision history} 
  gives a list of all changes to each version of the code.  The current review paper gives an overview of
  changes but is not meant to be complete. 
  
  \item A \href{http://www.nublado.org/wiki/FaqPage}{FAQ} page
  
  \item A summary of \href{https://www.nublado.org/wiki/CloudyOld}{old versions} of \Cloudy, 
  including links to download them.
  
  \item The \href{https://www.nublado.org/wiki/DeveloperPages}{developer pages} give links to
  our notes about developing \Cloudy.  You are most welcome to help!

\end{itemize}

\subsubsection{The \href{http://viewvc.nublado.org/index.cgi/?root=cloudy}{Subversion} repository}

The code, its documentation, test suite, and data 
live in a \href{http://subversion.apache.org/}{Subversion} 
repository.
The layout in the repository is conventional. 
The \textit{trunk} is the development version and is changed on a near-daily basis.

\textit{Branches} usually originate as copies of the trunk and can be separated into development
branches (to add new functionality) and release branches.
There is a C13 release branch which split off from the trunk in late 2012.  This branch is updated as
bugs are fixed but no new code development is done here.

\textit{Tags} are copies of a branch or trunk version.  
These do not change.  Released versions are tags.
For instance, the first release of C13 has the tag {\em C13.00}.

To assure the quality of the code, we run the test suite of the trunk on a
nightly basis provided there are changes. We also test the active release
branches and certain key development branches on a similar basis (albeit
somewhat less frequently). Additionally we test for common programming errors
such as array bounds violations and the use of uninitialized variables once or
twice a week. This way many errors can be caught quickly, preventing them from
causing problems in a release.

As an open source project, the entire repository is open to public view and download.
All versions of the code after the creation of the repository in late 2005 are available.
Older versions are maintained as separate tarballs on the 
\href{https://www.nublado.org/wiki/CloudyOld}{old versions} page.

\subsection{Testing}

The \Cloudy\ team has long participated in open comparisons of model predictions.
Such comparisons are a valuable way to exchange ideas and find problems, and
are the only way to validate projects as complicated as a modern spectral synthesis code
\citep{FerlandReliability01}.

Two meetings had been held by the time of F98 to compare predictions for ionized regions,
and a third was held soon after.
The first was organized by Daniel P{\'e}quignot in Paris in 1985 
\citep{Pequignot.D86Comparison-of-Photoionization-and-steady-shock}
but has no on-line proceedings available.
Two meetings were held in Lexington, the first a satellite of the STScI meeting
in honor of the 70$^{\rm th}$ birthdays of
Don Osterbrock and Mike Seaton \citep{1995aelm.conf.....W},
and a second as part of the Conference 
\textit{Spectroscopic Challenges of Photoionized Plasmas}
\citep{2001PASP..113.1024F}.
These comparisons are presented in
\citet{Ferland1995} and \citet{Pequignot2001} respectively.
Agreement at the 20\% - 30\% level for most important quantities was achieved
by \NEW{many} codes that participated in the workshops. 

A second form of testing is accomplished by running the code into well-posed physical limits.
Correct behavior in limiting cases gives some assurance that intermediate regimes are valid.
Examples include the Compton, LTE, molecular and low-density limits discussed in
Section \ref{sec:GridExtreme} below.
The code distribution includes a large test suite which exercises the code over its range of validity,
and includes embedded monitors that check that it obtains the expected result.
The test suite is designed so that the code can be automatically validated with little effort.

The scope of the simulations has been expanded to include atomic and molecular regions
in addition to ionized gas.
The goal is to calculate physical conditions of 
adjacent \hplus, \hone, and \htwo\ regions
(or \ion{H}{2} regions, PDRs, and molecular clouds) self-consistently.  
We begin a calculation at the face of a cloud illuminated by a hot O star 
and end in cold regions completely shielded from UV radiation \citep[see][]{Abel2005}.  
Such a calculation is a better representation of what actually goes on in nature, 
where \hplus, \hone, and \htwo\ regions are physically adjacent and the properties of each region 
depend on the radiative and dynamical coupling between the regions.  
This type of calculation is particularly advantageous in environments where the observed emission 
could come from more than one region.  

\Cloudy\ is well tested and in \NEW{good} agreement with other spectral synthesis codes that specialize in 
PDR modeling.  
The 2004 Leiden PDR meeting compared the results of several PDR codes for 8 benchmark calculations.  These calculations are summarized in \citet{Roellig2007}.  
These results show that \Cloudy\ agrees very well with the \hO/\htwo\ and \cplus/\cO/\CO\ transition, 
the dependence of other molecules with depth, temperature structure, and FIR emission-line spectrum.  
The \Cloudy\ test suite includes PDR calculations with parameters used by \citet{Tielens1985a}, 
\citet{Kaufman.M99Far-Infrared-and-Submillimeter-Emission-from}, 
and \citet{Le-Petit.F04H3-and-other-species-in-the-diffuse-cloud}.
These also agree fairly well.  
For more information, see the test suite input scripts that come with \Cloudy\ 
along with \citet{Abel2005} and \citet{Shaw2005}.

\citet{Abel2008} describe some differences between our predictions and
those of the PDR codes discussed in \citet{Roellig2007}.  These
largely are the result of our use of elementary processes rather than
fitting formulae in determining the physical state.  We show examples
of this below.

\subsection{Assessing the effects of uncertainties in atomic / molecular physics rates}
The atomic and molecular data set needed for a full simulation of the microphysics
of a non-equilibrium gas is vast, including ionization, dissociation, and recombination data for all species, 
along with internal energies, transition probabilities, and collision rates.
Many data are the results of theoretical calculations which are at the forefront of research
in computational atomic/molecular physics.
There will be gaps in these data, and in many cases, basic uncertainties.
\citet{Aggarwal.K13Assessment-of-atomic-data:-problems} review sources of these uncertainties
while \citet{Bautista.M13Uncertainties-in-Atomic-Data-and-Their}
look into how they propagate through spectral simulations,
both with emphasis on ions.
\citet{Wakelam.V10Sensitivity-analyses-of-dense-cloud} do a similar study
of molecular environments.

We have long included the ability to add a component of Monte Carlo Gaussian noise,
with a specified amplitude and FWHM, in our simulations.
We specify which component of the data to afflict with the noise, its amplitude, and dispersion. 
The data are altered when the code initializes and the disturbed data are used
throughout the calculation.
Each rerun of the simulation will have a different set of noise, 
as determined by randomly sampling the Gaussian distribution.
In many cases the noise is large - uncertainties can be as large as 0.2 - 0.5 dex.
The random numbers are Gaussian distributed in log space for this reason.

This capability is designed into \Cloudy\ to make it easy to examine the effects of uncertainties.
It has been applied in several studies.
\citet{Shaw2005} investigated the effects of uncertainties introduced by missing collisional rates
for \htwo.
\citet{Porter.R09Uncertainties-in-theoretical-HeI-emissivities:-HII-regions} and
\citet{Porter.R12Improved-He-I-emissivities-in-the-case-B-approximation}
documented how uncertainties in the photoionization cross sections, transition probabilities, and
collisional rates affect predictions of Case B \ion{He}{1} recombination coefficients.

Such studies make it possible to quantify how known uncertainties propagate into
the computed physical conditions or spectrum. 
It is important to remember that, in many cases, the dominant uncertainties are due to
physical processes which are not yet included.
Early simulations of \hii\ regions and planetary
nebulae failed because the importance of charge exchange and dielectronic recombination was
not understood.
As with any systematic error, 
the magnitude of the uncertainties can often only be known once they are removed.

\subsection{Modeling observations}
\label{optimize}

Observers are often faced with the problem that they have a set of
observations of a particular object and want to derive physical properties of
the object from these. The observations typically comprise spectral line and
continuum fluxes at various wavelengths and possibly other observables. From
these they would like to derive physical properties of the irradiating source
(e.g., the effective temperature) and/or the surrounding gas (e.g., the
density, electron temperature, chemical abundances, etc).

One way of achieving this is to ``reverse engineer'' the \Cloudy\ model by
assuming values for the physical parameters, calculating the model and
comparing the results to the observations. The quality of the fit is measured
by a $\chi^2$ value. The problem then reduces to finding the set of input
parameters that produce the best fit which is the lowest $\chi^2$ value. This
is a standard mathematical problem.

\Cloudy\ has an \textit{optimize} command that makes carrying out this task easy.
The heart of this command is the minimization algorithm for the $\chi^2$
function. There are two algorithms built in for doing this. The oldest one is
the SUBPLEX algorithm \citep{Rowan1990}. This is a generalization of the
well-known downhill simplex method AMOEBA. The second algorithm (which is the
default) is PHYMIR \citep{VanHoof1997} which was specifically designed for use
in \Cloudy. Both algorithms are robust against noisy functions which is a very
important feature since \Cloudy\ predictions are always noisy due to the use
of adaptive stepsize algorithms and finite precision iterative schemes.

The PHYMIR algorithm has two additional advantages that make it the preferred
method over the SUBPLEX algorithm. The first is that the PHYMIR optimization
process can be parallelized. If $N$ input parameters are varied then up to
$2N$ cores can be used simultaneously which can greatly speed up the
calculation. This is discussed in more detail in Sect.~\ref{parallel}.

The second advantage is that the PHYMIR algorithm periodically writes out
state files which can be used to restart an optimization run that failed
(e.g., due to a power failure) or ran into the maximum number of iterations
before the minimum was reached.

\subsection{Creating grids of calculations}
\label{sec:grid}

\begin{figure}
\centering
\includegraphics[width=\linewidth]{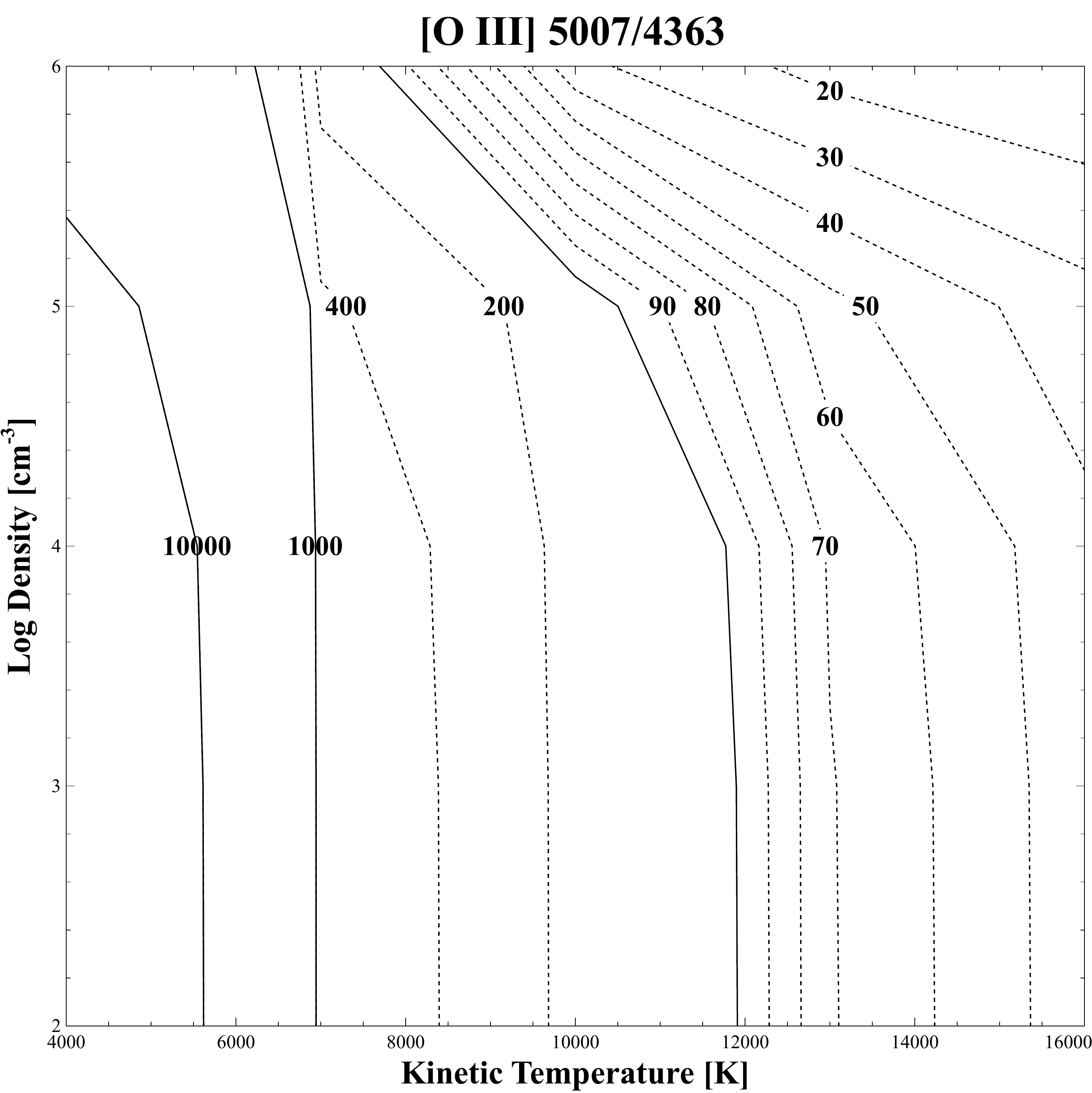}
\caption[Save line ratio example]{
\label{fig:SaveLineRatio}
The [O III] line $\lambda \lambda 5007 / 4363$ intensity ratio as a function of density and temperature.
}
\end{figure}

The \Cloudy\ \textit{grid} command, initially described by \citet{Porter2006},
makes it possible to vary input parameters to create large grids of calculations. 
Several parameters can be varied and the result of the calculation will be predictions 
for each of the grid points.
Figure~\ref{fig:SaveLineRatio} shows an example where a range of gas kinetic
temperature and density were computed and the [\oiii] $\lambda\lambda 5007, 4363$\AA\ lines
were saved.
Such diagrams can be used to deduce physical conditions in a cloud.

Predictions are usually saved with one of the \textit{save} commands described in \Hazy\
1. The predictions will normally be brought together into large files which
contain the output from all grid points.

\subsection{Cloudy on parallel computers}
\label{parallel}

Two \Cloudy\ commands can take advantage of multi-core computers and high
performance computing (HPC) clusters. These are the \textit{optimize} and {\it
  grid} commands described in Sects.~\ref{optimize} and \ref{sec:grid}. They run
\Cloudy\ as an ``embarrassingly parallel'' application, putting one model on
each CPU core. Run this way, the code can achieve nearly 100\% efficient use of
parallel computers.

The parallelization is implemented using two different techniques. The oldest
technique uses the \textit{fork} system call that is available under all UNIX
operating systems and Apple Darwin. The great advantages of this technique are
that no external libraries are needed (i.e.\ it works ``out of the box'') and
superior fault tolerance. All the work is done by the child processes, so even
in the unlikely event of a crash the parent process can continue, preventing
all work from being lost. The big disadvantage of this technique is that it
will only work on shared-memory machines so that it cannot take advantage of
modern HPC clusters. Currently only the \textit{optimize} command uses this
technique.

In view of the fact that HPC computing is moving away from shared-memory
machines towards large HPC clusters, we decided to redesign the parallel
infrastructure of the code. Since version C10.00 we support parallelization
under MPI version 2 or newer. Both the \textit{optimize} and \textit{grid} commands
can use this technique. The big advantage is that we can benefit from large
HPC clusters, which is important for large grid calculations which can now be
run as massively parallel applications. The disadvantage is that the user may
need to install an MPI environment or at the very least needs to get familiar
with MPI which unfortunately is not always as intuitive as we would like.

\subsection{Building complex geometries}
\label{sec:Cloudy3D}

Clumping can be included and complex source geometries can be simulated.  
There are several general considerations.

There are powerful selection effects governing the formation of emission lines when a range of densities
exist.  You will tend to observe the highest-density regions because the
emission per unit volume is proportional to the square density if the line is below its
critical density (AGN3 Section 3.5).
Only with a fine-tuned mix of
densities, where the volume of material at each density exactly compensates
for the change in emissivity, will an observer notice emission from a range
of densities.  Claims that a range of
densities contribute to a single emission line should be met with some skepticism.
This would require an amazing coincidence 
\citep{Ferland.G11Molecular-hydrogen-in-NLSys-and-its-implications}.

But clumps do exist.
If the clump size is small compared with the
physical thickness of the region then they can be treated with a filling
factor (see \citet{Osterbrock1959}
and AGN3 Section 5.9).
In this
case the gas is modeled as small clumps that are surrounded by vacuum or
much lower-density gas.
This is done by simply including the
\textit{filling factor} command to specify the fraction
of the volume that is filled by clumps.

If the clumps are larger than the physical thickness of the \NEW{line-forming} region
then each clump will have its own ionization structure.
This is the ``LOC''
model of quasar emission-line clouds described by
\citet{Baldwin1995} and \citet{FergusonKoristaBaldwin1997}.
The model is developed in several papers by the same team.
In this case
we compute grids of models and save the results.  The spectra are then
co-added using distribution functions to describe the range of cloud
properties.  The final spectrum depends on these distribution functions.
\citet{GiammancoEtAl_clumpsCloudy04} show \Cloudy\ calculations
where optically thick clumps are present in the ISM.

In practice we normally use the \emph{grid} command (S\ref{sec:grid}) for this, but there are
circumstances where complex changes in parameters may be needed.
The program \cdFilename{mpi.cpp} in the programs directory in the code distribution
computes a grid of models and extracts the predictions using MPI on a
distributed memory machine.

\begin{figure*}
\centering
\includegraphics[width=0.7\linewidth]{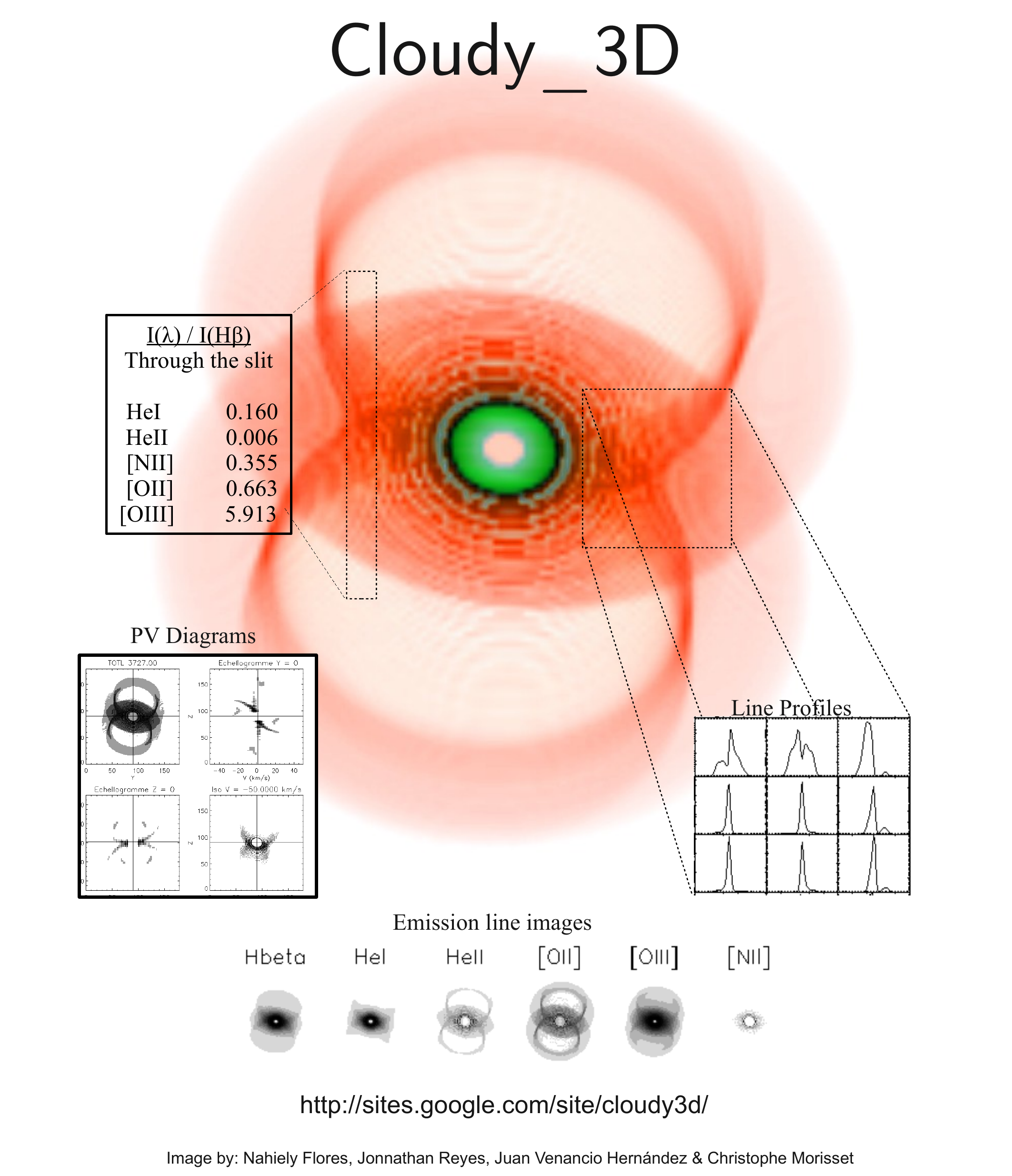}
\caption[Cloudy\_3D simulation of a planetary nebula]
{A 3-color image of a Hourglass-type nebula, 
obtained by running \textsc{Cloudy\_3D} \citep{MorissetCloudy3D06}.
Colors are [\nii ] (orange) and [\oiii ] (green) emission. 
Emission line profiles are shown for [\nii ] lines. 
Intensities through any given slit can be obtained. 
Position-velocity diagrams are obtained as well as
channel maps, for any line. 
Emission line surface brightness maps are
also available for any line computed by \Cloudy. 
Statistical tools to
analyze emission-line properties are also provided.
}
\label{fig:Cloudy3D}
\end{figure*}

Another approach is for a driving program to use \Cloudy\ to compute differential volume elements of
a large and complex structure, and then integrate to get the next emission.
An example is the \textsc{Cloudy\_3D} code\footnote{%
  \url{http://sites.google.com/site/cloudy3d/}.
}
described in \citet{MorissetCloudy3D06}
and \citet{MorissetStasinskaCloudy3D08}.
\textsc{Cloudy\_3D} was used to compute
the image shown in Figure~\ref{fig:Cloudy3D}. 
The more recent \textsc{pycloudy} code by the same author\footnote{%
  \url{https://sites.google.com/site/pycloudy/home}.
}
is a more general tool for controlling and analyzing 
multiple \Cloudy{} runs via scripts.
The \textsc{Rainy3D} code is another example 
\citep{MoraesDiazRAINY09}.

\subsection{Spectral energy distributions from stellar atmosphere grids}
\label{sec:StellarAtmospheres}

The heart of any photoionization simulation is the SED of the incident
radiation field.
It is this energy which is reprocessed by the cloud to produce the observed nebular emission.
Several groups have created large grids of stellar SEDs using advanced stellar atmosphere codes.
Other groups have used these data to create stellar population synthesis models that give
the integrated spectrum of a galaxy as a function of time after a starburst.
\Cloudy\ can  interpolate on SED grids having an arbitrary
number of dimensions (these might include surface temperature, gravity, chemical composition, mass loss rate, age, etc)
and include this in the incident radiation field.

\begin{figure}
\centering
\includegraphics[width=\linewidth]{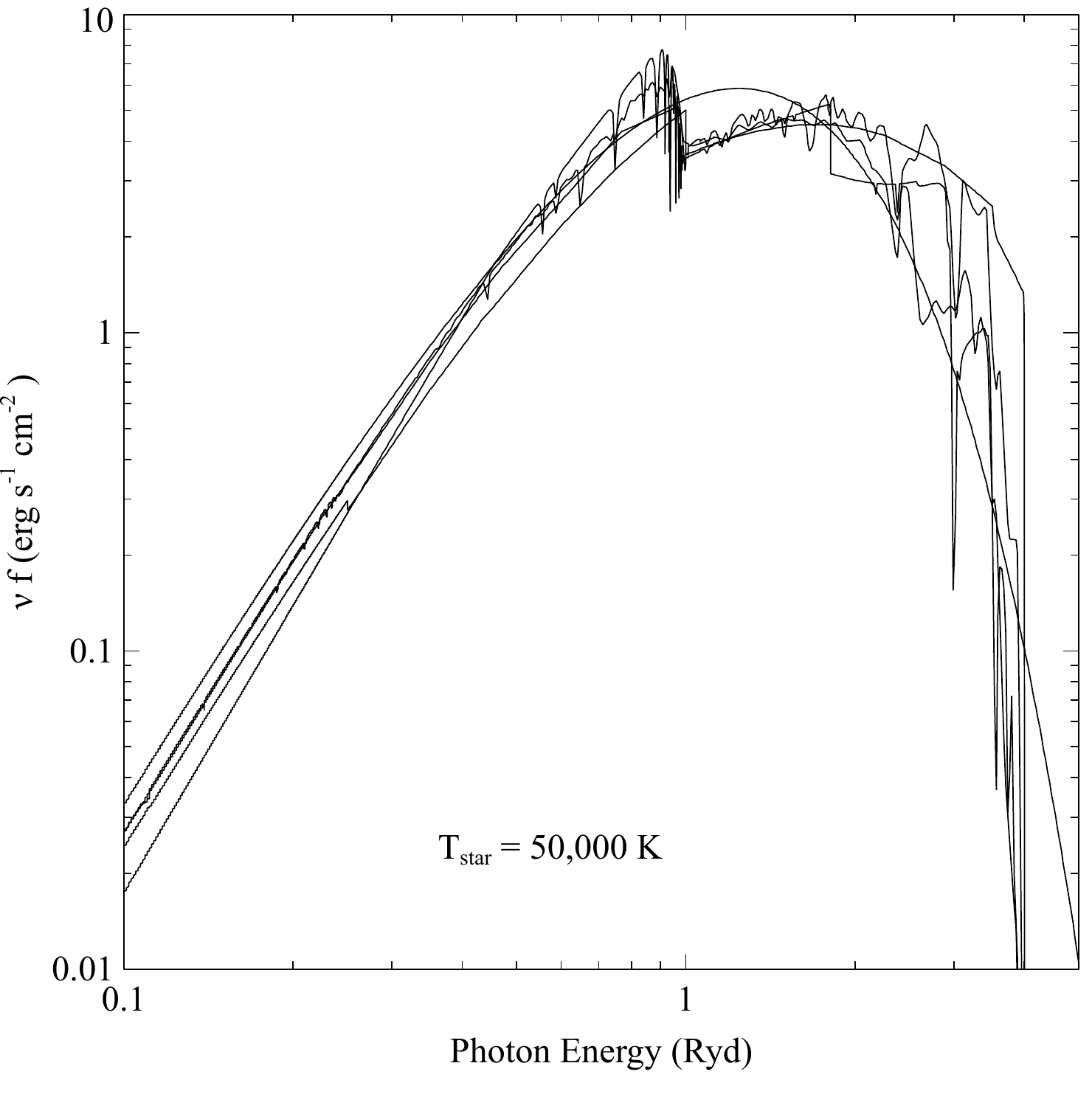}
\caption[Stellar radiation fields]
{\label{fig:StarsContinuum}This figure shows the emergent
radiation field predicted by five 5\e4 
K stars included with the code.  The smoothest is the blackbody, and the
\citet{Kurucz1991} and \citet{Rauch1997} atmospheres show the most structure.}
\end{figure}

Figure~\ref{fig:StarsContinuum} compares predictions for
five of the 5\e 4 K SEDs that are available.  
These include a blackbody and atmospheres computed by \citet{Mihalas1972}, \citet{Kurucz1979}, \citet{Kurucz1991} and \citet{Rauch2003}.
All were normalized
to have the same total luminosity ($10^{38}$ erg s$^{-1}$)
observed from a distance of $10^{18}$~cm.
Note the order of magnitude dispersion among the continua for
energies around 4 Ryd.
This can have a major effect on the \Cloudy\ modeling results, showing the crucial
role that the stellar SED plays.

Numerous stellar grids can be used with little additional work.
In some cases the SED data are stored on the author's web site while
in others they are stored on the \Cloudy\ web site. A convenient page
providing links to all the necessary files can be found at
\href{http://wiki.nublado.org/wiki/StellarAtmospheres}{wiki.nublado.org/wiki/StellarAtmospheres}.
This web page also gives more computational details and links to the papers describing the grids.

These are the most important SED grids currently supported by \Cloudy:
\begin{enumerate}
\item
The Atlas grids. There are two versions of these, the preferred one being the
new-ODF grids described in \citet{Castelli2004}. The older generation of Atlas
model atmospheres described in \citet{Kurucz1991} is also supported. These
grids can be useful if you need models for extreme metalicities not covered by
the new-ODF grids. Both grids contain LTE, plane-parallel, hydrostatic model
atmospheres with effective temperatures ranging between 3\,500 and 50\,000~K.
\item
The Tlusty OSTAR2002 and BSTAR2006 grids described in \citet{Lanz2003} and
\citet{Lanz2007}. These grids contain non-LTE, line-blanketed, plane-parallel,
hydrostatic O and B star SEDs. We also support merged OSTAR2002/BSTAR2006
grids covering a temperature range between 15\,000 and 55\,000~K.
\item
The WMbasic and CoStar O and B star grids. These are two small grids of
non-LTE, line-blanketed, and wind-blanketed models. The first grid is described
in \citet{Pauldrach2001} and the second in \citet{Schaerer1997}.
\item
All PN central star SED grids computed by T. Rauch. These include the H-Ni, PG
1159, and C/O white dwarf grids, as well as the pure hydrogen, pure helium and
H+He grids. The older H-Ca grids are also supported, though for most purposes
they have been superseded by the H-Ni grids (unless you need models with
T$_{\rm eff} >$ 190\,000~K). All grids contain non-LTE, line-blanketed,
plane-parallel, hydrostatic model atmospheres. They are described in
\citet{Rauch1997} and \citet{Rauch2003}. The temperature range typically is
between 50\,000 and 190\,000~K, though some grids have a different range.
\item
Stellar population synthesis models from Starburst99 \citep{Leitherer1999} and
PopStar \citep{Molla2009}. Typically the user will do their own run and
generate a \Cloudy\ grid from the output using either \Cloudy\ commands
(Starburst99) or a script supplied on the \Cloudy\ web site (PopStar).
\end{enumerate}

Many of the grids are very large and accessing them as ASCII files would be slow. 
They are ``compiled'' to create direct access binary files as part of the installation procedure.
Once complete the stellar SEDs can then be accessed with the appropriate command
in the simulation control deck.
Logarithmic interpolation is done to create model atmospheres with any set of specified parameters
using nearby models from the original grid.

The code is very flexible and allows users to create their own SED grids, e.g.,
from a Starburst99 run. As a result we can also add support for new grids
during a release cycle when the need arises. This is possible because no code
changes are needed to do this.

\subsection{Citing \Cloudy\ and its underlying databases}

\Cloudy\ is a research project that involves the creative efforts of many people.  
When used in publications it should be cited as follows:
``\textit{Calculations were performed with version C13.00 of \Cloudy, 
last described by Ferland et al. (2013)}'', where this paper is the reference.
The specific version of the code, written as C13.00 in this example,
should be given so that, in case any future questions arise,
it will be possible to reproduce the calculation using the archived versions
on \href{http://www.nublado.org}{www.nublado.org}. 

We are now moving the atomic and molecular data to external databases.
These are replacing our internal database, which had been embedded in the source.
Many recombination coefficients are based on 
\citet{BadnellEtAl03} and  \citet{Badnell06} and posted on
the web sites \url{http://amdpp.phys.strath.ac.uk/tamoc/RR/} and \url{http://amdpp.phys.strath.ac.uk/tamoc/DR/}.
Much of the molecular emission data is from LAMDA 
\citep{Schoier.F05An-atomic-and-molecular-database-for-analysis} as accessed on Dec. 18, 2010,
as well as the JPL \citep{Pickett1998} and CDMS \citep{Mueller2001, Mueller2005} databases.
Much of the ionic emission data is from CHIANTI,
as described by 
\citet{Dere.K97CHIANTI---an-atomic-database-for-emission} and
\citet{Landi2012}, using version 7.0.
Much of the \htwo\ data is from \citet{Wrathmall2007},
\citet{Abgrall.H94The-Bprime1Sigmau-rarr-X1Sigmag-and-D1Piu},
and the Meudon web site
(\url{http://molat.obspm.fr/index.php?page=pages/Molecules/H2/H2can94.php}). 

All of these databases play a major role in most calculations.  
We ask that users cite both \Cloudy\ itself, and those underlying databases,
in any publications.
These databases can only thrive if their role is properly acknowledged.
 We provide a \textit{print citations} command that will
provide the correct citations in a format that can be easily copied and pasted into papers.

\section{Applications}

\subsection{XDRs}

\begin{figure}
\centering
\includegraphics[width=\linewidth]{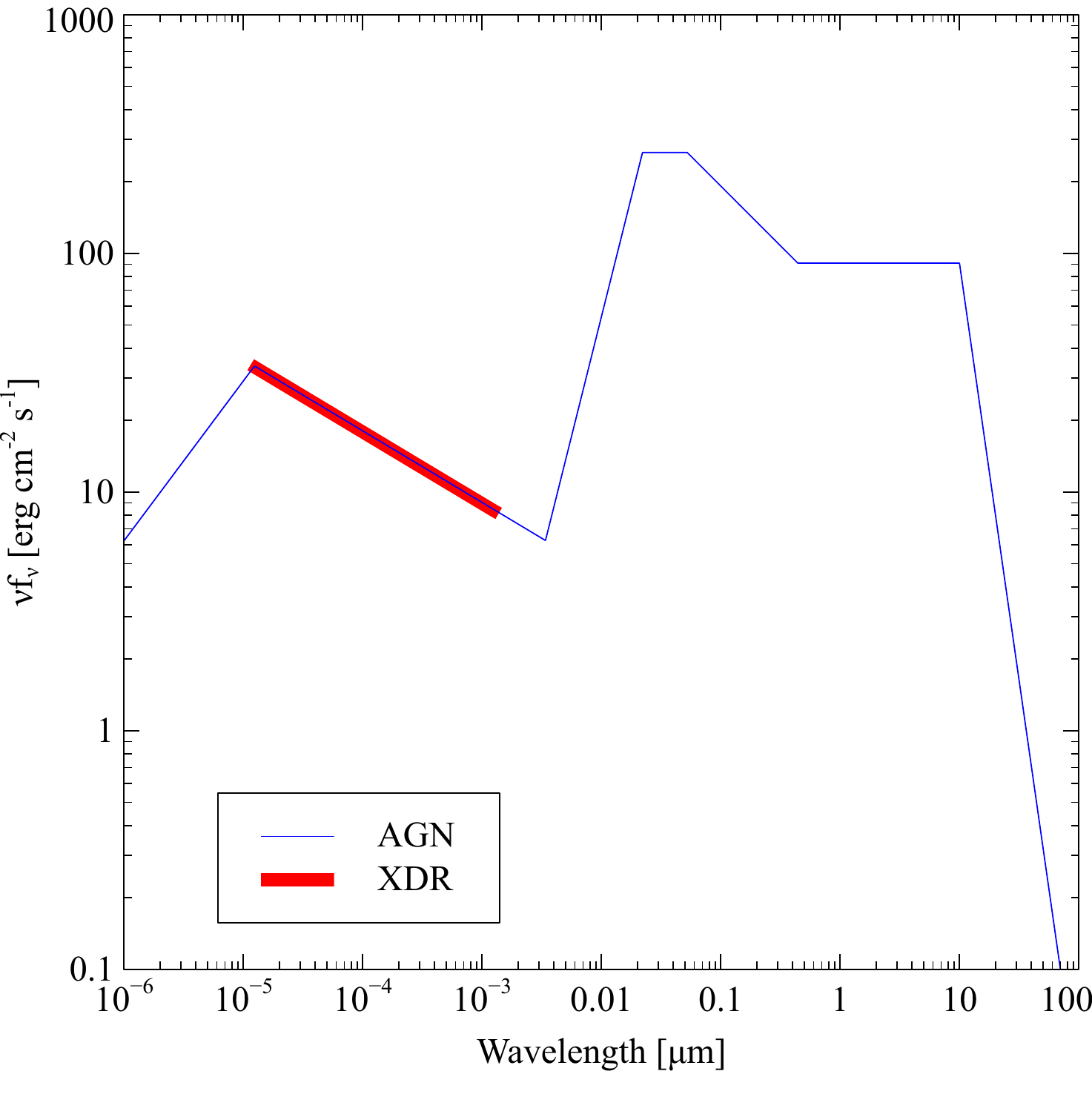}
\caption[XDR and AGN continua]{
\label{fig:XdrAgnSed}
The thin line shows a mean AGN SED as deduced by \citet{Mathews1987}.
The thick line shows the truncated SED considered in XDR calculations.
Both SEDs are built into \Cloudy\ and were normalized to have the same X-ray flux.}
\end{figure}

The term X-ray Dissociation Region or ``XDR'' was coined by \citet{Maloney1996} to 
describe atomic regions near X-ray sources.
(A somewhat similar calculation had been presented by \citet{Ferland1994}
in the context of optical filaments in cool-core clusters of galaxies.)
In keeping with traditions established in the study of PDRs, a truncated SED, 
including only photons between 1 -- 100 \keV, was considered.
Figure~\ref{fig:XdrAgnSed} shows the \citet{Maloney1996} X-ray continuum together with the mean
AGN SED derived by \citet{Mathews1987}.
Both SEDs are built into \Cloudy.

A second Leiden meeting on radiatively excited atomic and molecular regions was held
in 2012\footnote{\url{http://www.lorentzcenter.nl/lc/web/2012/482/info.php3?wsid=482}}.
The meeting considered five PDR and four XDR simulations.
The web site\footnote{\url{http://home.strw.leidenuniv.nl/~loenen/LC-CO/}} 
gives some details along with results of the participating codes.
We agreed with the PDR results, to within the considerable scatter,
as had been found by \citet{Roellig2007} and \citet{Abel2008}.
\NEW{However we systematically found less CO in the XDR simulations, 
as shown in the plots posted on the web site.
We eventually traced this down to the cloud thickness which had not been
specified for the XDR sims.
Marcus R{\"o}llig kindly provided us with results submitted by other participants,
and we have recomputed the XDR sims with a total hydrogen column density
of $N(\textrm{H}) = 3\times 10^{24} \pscm$.}

\NEW{In this section we consider some details of our treatment of XDRs,
since we have never directly considered such simulations before.
Although our final results are within the substantial scatter of the results
presented at the meeting, there are some interesting aspects of the
calculation which we discuss next.}

\NEW{Here we consider the XDR2 test in some detail.}
This simulation has a hydrogen density of $n_{\rm H} = 10^3 \pcc$, the hydrogen column
density given above (corresponding to point-source  \Av\ $\sim 10^3$ mag) 
and an X-ray flux of 270 \ergpscmps.
The gas ionization is proportional to the dimensionless ionization parameter
\begin{equation}
U = \frac{\phi({\rm H})}{c \ n({\rm H})}
\end{equation}
where $\phi({\rm H})$ is the flux of hydrogen ionizing photons \citep{AGN3}.
This simulation has the highest ionization parameter of the XDR tests,
\NEW{and so is one where our detailed treatment of singly and doubly charged ions makes a difference.}

\begin{figure}[!tbp]
\centering
\includegraphics[width=\linewidth]{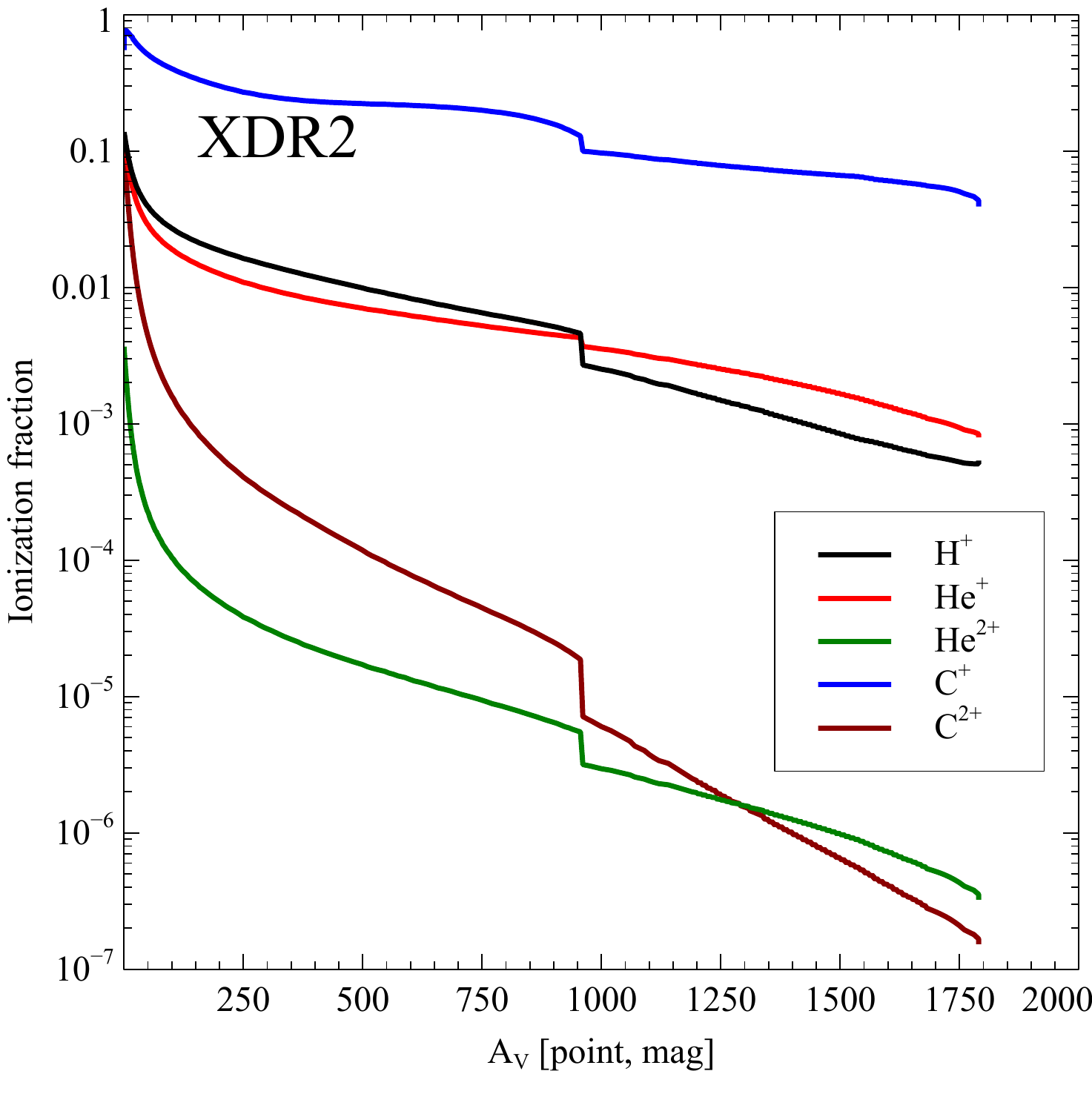}
\caption[Ionization fractions for H, He, and C]{
\label{fig:XDR2Ionization}
The ionization fractions for H, He, and C are shown as a function of depth,
expressed as the point-source \Av.}
\end{figure}

Photoionization by the incident radiation field, and by diffuse EUV line emission,
emission lines produced by the XDR gas,
produces a moderate level of ionization throughout the XDR2 cloud.
Figure~\ref{fig:XDR2Ionization} shows the ionization fractions for H, He, and C
as a function of the point-source \Av.
There are significant amounts of doubly ionized species.
The most important of the ions shown is
\heplus, which destroys CO by charge exchange.

\begin{figure}
\centering
\includegraphics[width=\linewidth]{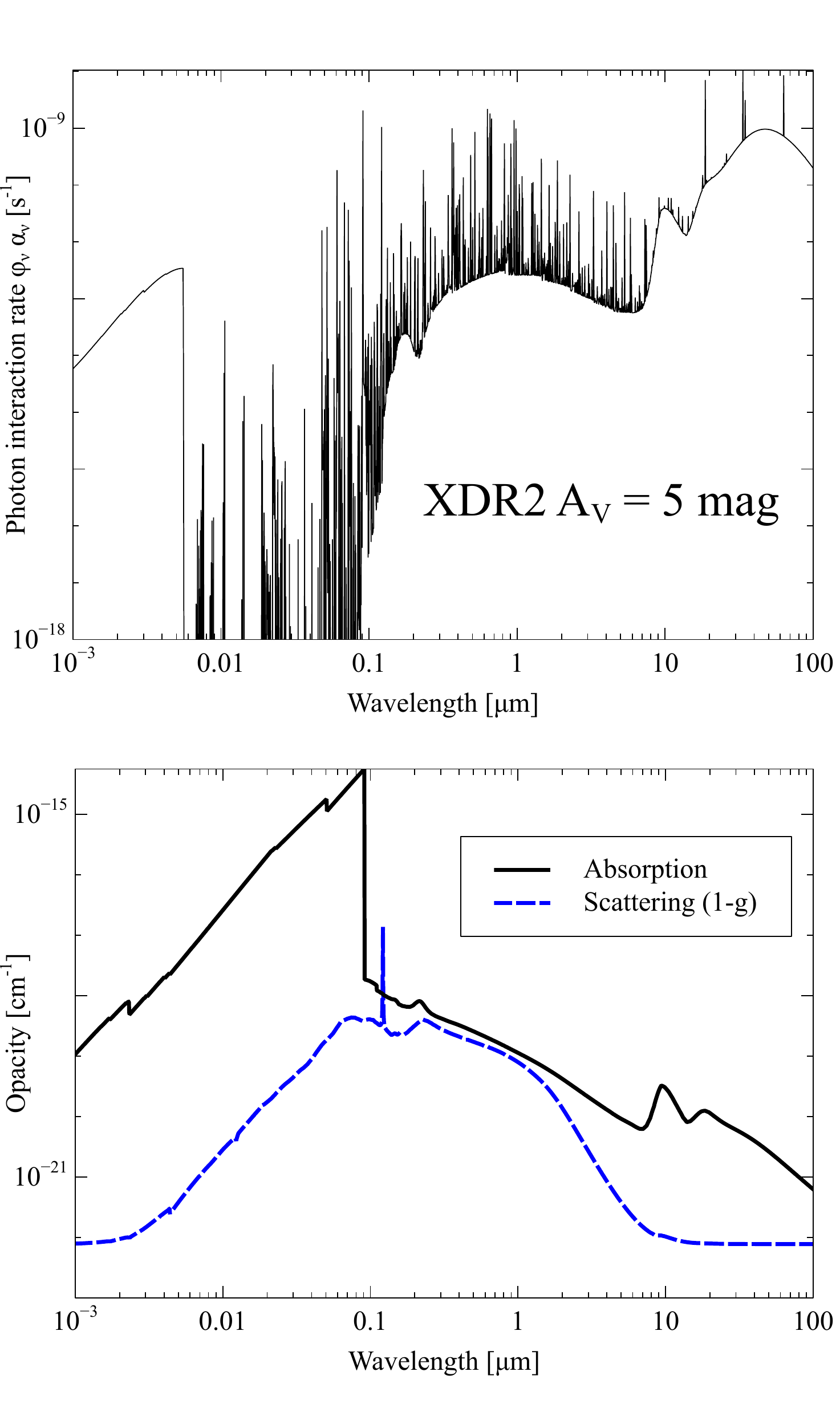}
\caption[Mid-plane radiation field in XDR2]{
\label{fig:XDR2midplaneRadField}
Components of the radiation field at a depth corresponding to  \Av\ = 5 mag in the XDR2 simulation.
The lower panel shows the absorption and scattering opacities while the
upper panel shows the photon interaction rate.}
\end{figure}

\NEW{Figure~\ref{fig:XDR2midplaneRadField} shows a representation of the 
internal radiation field at a depth corresponding to \Av\ = 5 mag.
Although this is a shallow depth for an XDR, it is also where the warm gas that
is most efficient in producing emission is found.}
The lower panel shows the absorption and scattering opacities,
where the latter includes the factor $(1-g)$ which discounts forward scattering.
Grains are the major contributor to the total scattering across the UV and optical,
with Thomson scattering being dominant at 
the shortest and longest wavelengths.
\ion{H}{1} Rayleigh scattering is
responsible for the feature at $\sim 0.1216 \micron$.

Grains are the dominant absorption opacity source across the UV, optical, and IR, 
with several resonant features visible.
Below 0.0912 \micron\ gas opacity dominates, with the strongest edge due to \hO.
Edges due to He and inner shells of the heavy elements are visible at shorter wavelengths.

The upper panel shows the local photon interaction rate, $\phi_{\nu} \alpha_{\nu}$, 
where $\alpha_{\nu}$ is the gas opacity [\pscm].
The local photon flux $\phi_{\nu} = 4 \pi \bar{J}/h\nu$ [\pscmps] includes all components of the 
\NEW{radiation field at that point,}
including the attenuated incident SED and the local diffuse line and continuous emission.
The strong lines in the FUV and EUV\footnote{\NEW{We follow standard astronomical nomenclature and 
refer to the region $6 - 13.6 \eV$ (912\A{} to 2000\A) as FUV:
with EUV the region $13.6 - 56.4 \eV$ (or 228\A\ to 912\A)
and  XUV $56.4 \eV -$ few hundred \eV\ ($\lambda < 228\A$).}}
 are the result of the solution of the many-level 
iso-sequence atoms as described in previous sections, 
and have intensities \NEW{that are} fully self consistent with the opacities shown in the lower panel, 
the level of ionization, gas temperature, and optical depths.
There are significant sources of ionizing radiation in addition to the attenuated incident XDR continuum.
The most important are EUV recombination lines of \ion{He}{1} and \ion{He}{2}.
Direct photoionization by the incident continuum produces a trace amount of \hePP\
while the EUV emission lines, together with the incident continuum,
produce a moderate amount of \heplus\ and \hplus.

\begin{figure}
\centering
\includegraphics[width=\linewidth]{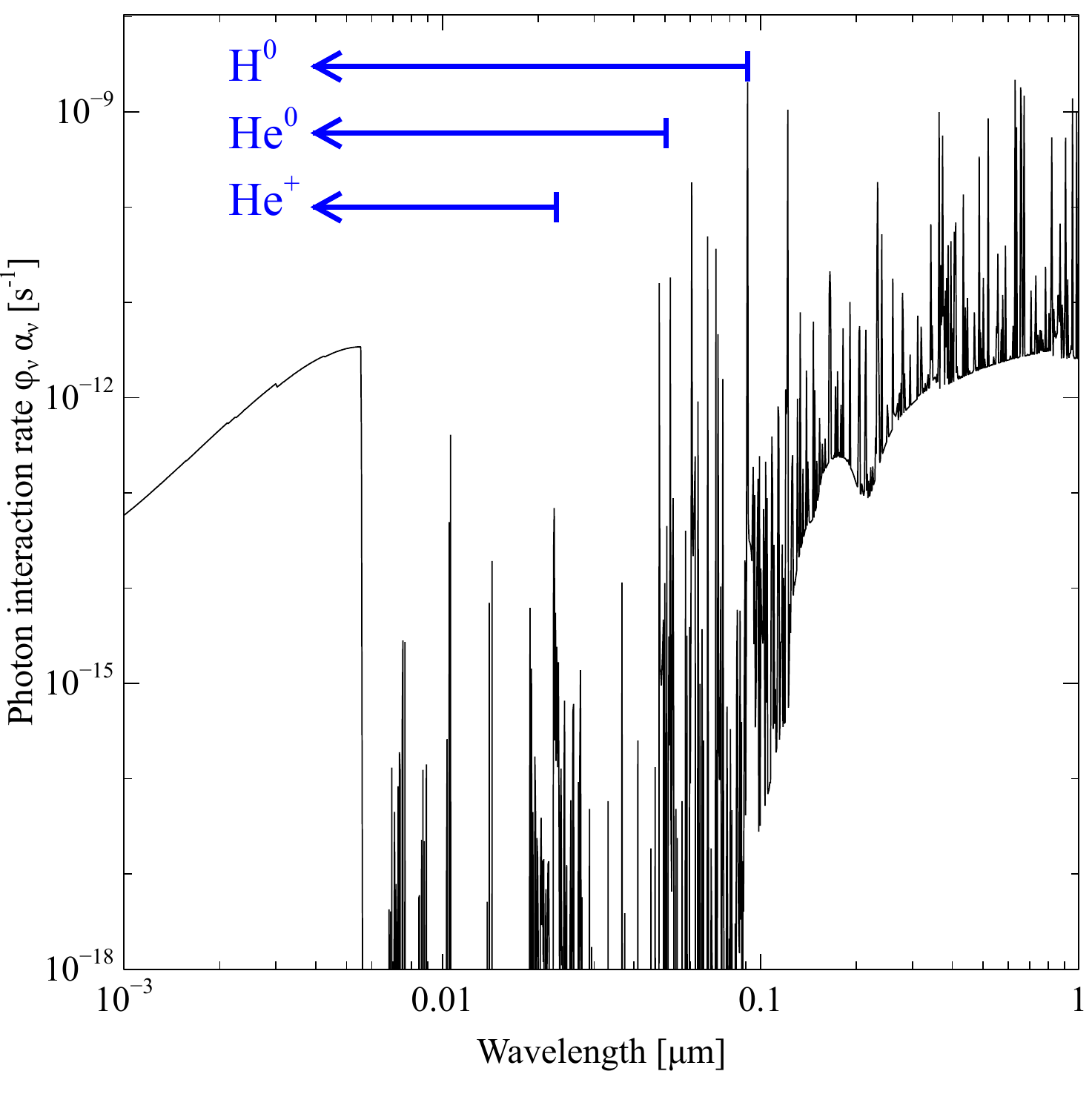}
\caption[Mid-plane radiation field at EUV wavelengths]{
\label{fig:XDR2midplaneZoom}
This shows a small portion of the radiation field around wavelengths capable
of ionizing H and He.
The horizontal lines indicate thresholds for photoionization of \hO, \heo, and \heplus.}
\end{figure}

Figure~\ref{fig:XDR2midplaneZoom} shows a zoom into the photon interaction rate 
$\phi_{\nu} \alpha_{\nu}$ within the hydrogen-ionizing radiation field.
The total photoionization rate of a species is the integral of  
$\phi_{\nu}$ over the photoionization cross section $\sigma_{\nu}$ (AGN3),
while the Figure shows the total interaction rate, evaluated using the computed total opacities.
The horizontal lines indicate the range of wavelengths which can ionize H and He.
For reference, photoionization cross sections fall off with decreasing wavelength as a
power law ranging from $\sigma_{\nu} \sim \lambda^{-3}$ for \hO\ and \heplus\ to 
$\sigma_{\nu} \sim \lambda^{-1}$ for \heo.
The attenuated incident XDR continuum is the dominant contributor to the \heplus\ 
photoionization rate.
Recombination lines of \ion{He}{2}, with a significant contribution from the XDR continuum,
dominate photoionization of \heo, and these lines, together with \ion{He}{I} lines,
ionize \hO.

The cumulative effect is that a significant amount of \heplus\ exists across \NEW{shallow parts} of the cloud
(Figure~\ref{fig:XDR2Ionization}).
Charge transfer between \heplus\ and CO is the dominant CO destruction process
in regions that are well-shielded from FUV radiation:
\begin{equation}
{\rm He}^+  + {\rm{CO}} \to {\rm He} + {\rm C}^+ + {\rm O}
\end{equation}
\citep{Anicich.V77Product-distributions-for-some-thermal,Laudenslager.J74Near-thermal-energy-charge}.
The large amount of \heplus\ results in efficient destruction of CO,
and little CO exists as a result.
The strong \ion{H}{I}, \ion{He}{1}, and \ion{He}{2} recombination lines
heat the gas through both direct photoionization, and grain electron ejection.

\begin{figure}
\centering
\includegraphics[width=\linewidth]{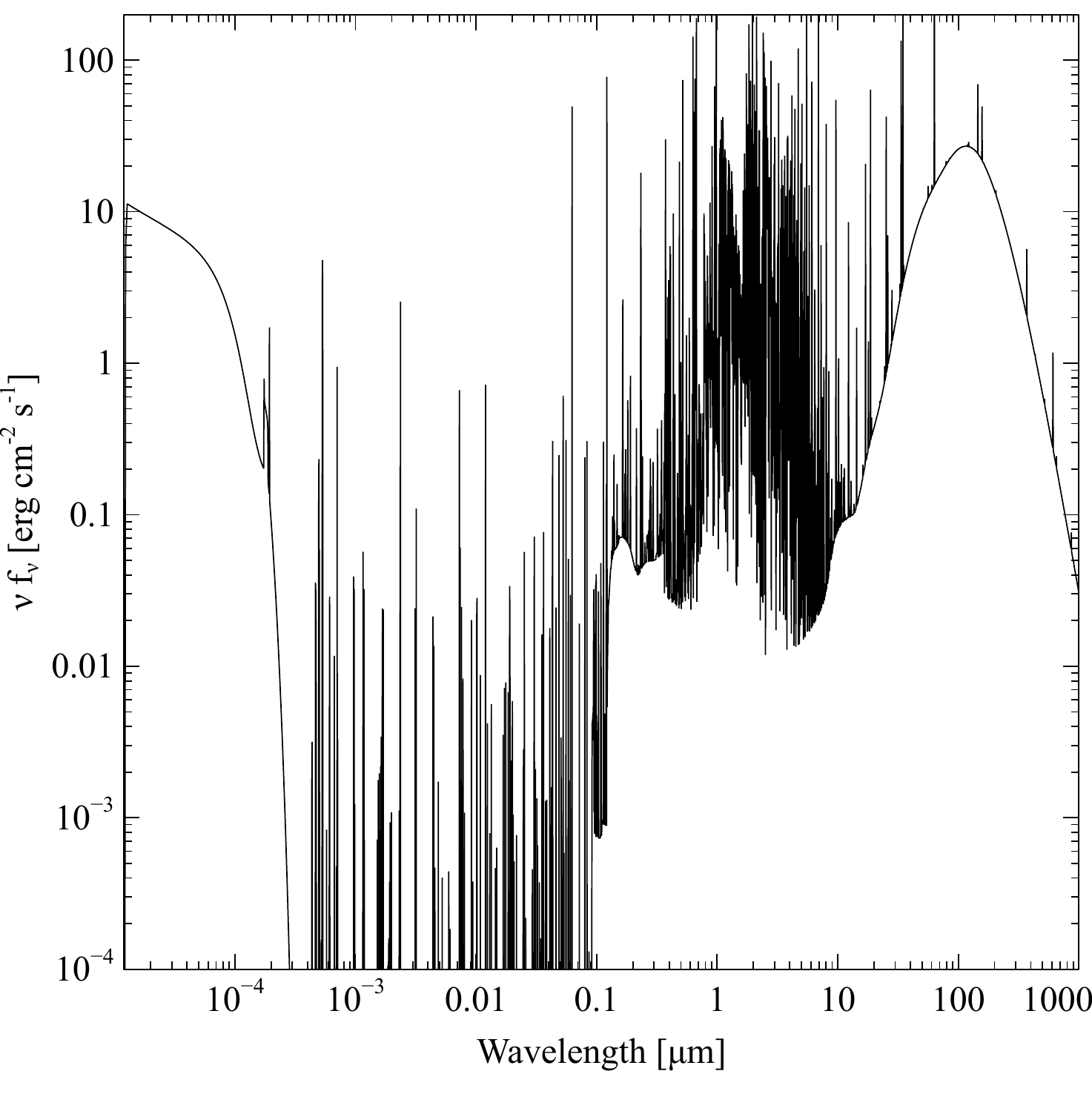}
\caption[The spectrum emergent from the XDR2 cloud]{
\label{fig:Xdr2Con}
\NEW{The spectrum emergent from the XDR2 cloud.
Thermal dust emission dominates in the IR while the incident XDR continuum dominates in the X-ray.
A rich UV, FUX, and EUV spectrum is emitted by atoms and ions within the gas.}}
\end{figure}

\NEW{Figure~\ref{fig:Xdr2Con} shows the spectrum emitted by the XDR2 cloud.
The thermal infrared continuum emitted by grains, with the silicate
10 \micron\ feature in emission, is apparent.
(PAHs were not included so their features are absent.)
The single strongest line is \hi\  \la, produced by ionized gas present within the cloud. 
There is a significant amount of 
EUV
emission at $\lambda < 912$\AA.
The blended (at this scale) cluster of lines between $1 - 10\  \micron$ is
mainly produced by \htwo.}

\begin{figure}
\centering
\includegraphics[width=\linewidth]{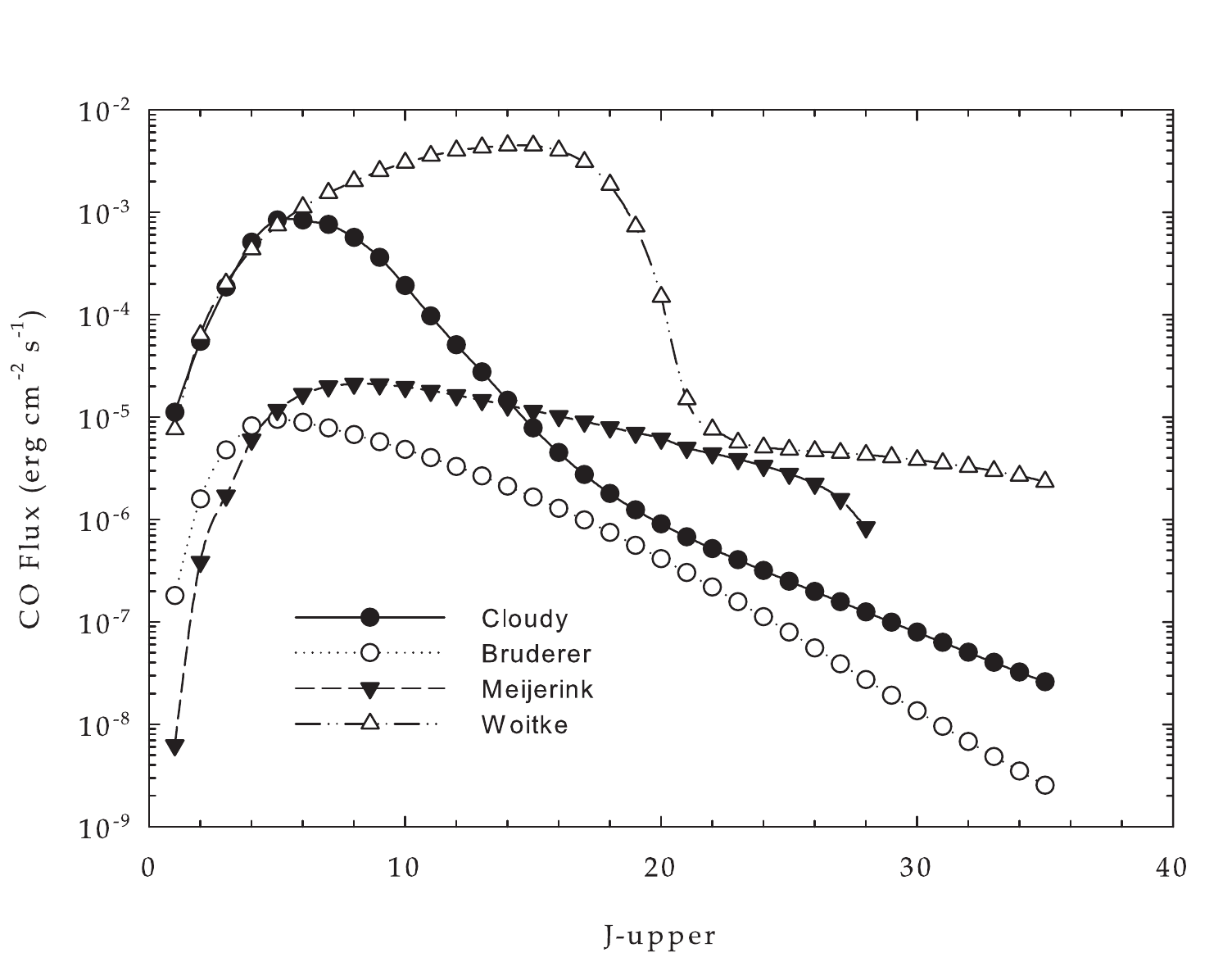}
\caption[Predicted CO 1-0 intensity for Leiden 2012 XDR models]{
\label{fig:XdrCoIntensity}
\NEW{The XDR2 CO rotation ladder predicted by codes represented at the Leiden 2012 meeting.
The meeting web site gives more details.}}
\end{figure}

\NEW{The goal of the 2012 Leiden workshop was to compare predictions of the CO rotation ladder.
Figure \ref{fig:XdrCoIntensity} shows our predictions together with 
those of other workers, kindly provided by Marcus R{\"o}llig.
Our predictions lie within range of results given by other codes,
as we had previously found for PDRs \citet{Roellig2007}.}

Table~\ref{tab:XDR2lines} gives the intensities of lines predicted to have
emergent intensities brighter than 10\% of the 
[\ion{C}{2}] $\lambda 158 \micron$ line in the format used by \Cloudy.
We intentionally present Table~\ref{tab:XDR2lines} in the format used by the
code, as an introduction to its output.
Each line is indicated by a label in the \Cloudy\ output, 
given as the first column in the Table,
and a wavelength, given in the second column.
The line label uses the compact notation ``\verb|C  2|'' for [\ion{C}{2}] so the
label is intended for identification rather than spectroscopic notation.
In the \Cloudy\ output, as in Table~\ref{tab:XDR2lines}, ``\texttt{A}'' and ``\texttt{m}'' indicate \AA\ and \micron\
respectively.
A very large number of lines are predicted by \Cloudy.
We provide a \textit{save line labels} command which will create
a list of all emission lines with labels, wavelengths, and a comment indicating the line's origin.
There is a discussion of the various line entries in Part 2 of \Hazy, the code's documentation.

The third column gives the intensity, $4 \pi J({\rm line})$ [\ergpscmps] of each line.  This is the total
emission radiated into $4\pi \sr$ from a unit area of cloud,

\begin{table}[t!]
\centering
\caption{\NEW{Line intensities for XDR2 simulation}}
\label{tab:XDR2lines} 
\small
\tablecols{4}
\setlength\tabnotewidth{\linewidth}
\setlength\tabcolsep{2\tabcolsep}
\begin{tabular}{@{} cr@{\hspace*{2em}} cr @{\quad\quad}}
\toprule
Spectrum
&
\multicolumn{1}{@{}c}{
Wavelength 
} 
& 
Intensity\tabnotemark{a} 
& 
\multicolumn{1}{@{}r@{}}{
\(I /[\ion{C}{2}]\ \lambda 158 \micron\)
}
\\ \midrule
\verb|FeKa| & \verb|1.78A|  & \(5.07\e{-1}\) & 5.16\phn \\
\verb|H  1| & \verb|1216A|  & \(1.73\e{-1}\) & 1.76\phn \\
\verb|C  2| & \verb|2326A|  & \(1.29\e{-1}\) & 1.32\phn \\
\verb|O  2| & \verb|3727A|  & \(1.34\e{-2}\) & 0.137 \\
\verb|O II| & \verb|3729A|  & \(1.63\e{-1}\) & 1.66\phn \\
\verb|S  2| & \verb|4074A|  & \(7.55\e{-2}\) & 0.768 \\
\verb|N  1| & \verb|5199A|  & \(1.66\e{-2}\) & 0.169 \\
\verb|O  1| & \verb|6300A|  & \(4.57\e{-2}\) & 0.464 \\
\verb|O  1| & \verb|6363A|  & \(1.48\e{-2}\) & 0.151 \\
\verb|H  1| & \verb|6563A|  & \(2.06\e{-2}\) & 0.209 \\
\verb|H-CT| & \verb|6563A|  & \(5.55\e{-3}\) & 0.056 \\
\verb|N  2| & \verb|6584A|  & \(2.16\e{-2}\) & 0.219 \\
\verb|S II| & \verb|6716A|  & \(1.77\e{-2}\) & 0.181 \\
\verb|S II| & \verb|6731A|  & \(1.43\e{-2}\) & 0.145 \\
\verb|S  3| & \verb|9069A|  & \(1.24\e{-2}\) & 0.126 \\
\verb|S  3| & \verb|9532A|  & \(3.22\e{-2}\) & 0.327 \\
\verb|C  1| & \verb|9850A|  & \(3.32\e{-2}\) & 0.338 \\
\verb|S II| & \verb|1.029m| & \(1.34\e{-2}\) & 0.136 \\
\verb|S II| & \verb|1.032m| & \(1.83\e{-2}\) & 0.186 \\
\verb|S II| & \verb|1.034m| & \(1.30\e{-2}\) & 0.132 \\
\verb|He 1| & \verb|1.083m| & \(2.34\e{-2}\) & 0.238 \\
\verb|S  3| & \verb|18.67m| & \(9.64\e{-2}\) & 0.980 \\
\verb|S  3| & \verb|33.47m| & \(3.37\e{-1}\) & 3.42\phn \\
\verb|Si 2| & \verb|34.81m| & \(1.50\e{-1}\) & 1.52\phn \\
\verb|O  1| & \verb|63.17m| & \(6.04\e{-1}\) & 6.14\phn \\
\verb|O  1| & \verb|145.5m| & \(1.48\e{-1}\) & 1.50\phn \\
\verb|C  2| & \verb|157.6m| & \(9.84\e{-2}\) & 1.00\phn \\
\verb|C  1| & \verb|369.7m| & \(1.40\e{-2}\) & 0.142 \\
\verb|TIR | & \verb|1800m|  & \(1.54\e{-1}\) & 1.56\phn \\
\verb|TALL| & \verb|10000A| & \(4.83\e{+1}\) & 491.\phn\phn\phn \\
\bottomrule \\
\tabnotetext{a}{Intensity $4\pi J(\mathrm{line})$ with units \ergpscmps }
\end{tabular}
\end{table}

Several lines deserve special mention.
``FeKa 1.78A'' is the Fe K$\alpha$ X-ray line and is mainly 
produced by Fe locked in silicate grains.
The \ha\ line is predicted to have a significant contribution by
mutual neutralization excitation, indicated by the label ``H-CT''.
This is the process
\begin{equation}
{\rm H}^-  + {\rm{p}} \to {\rm H}^*(n=3) + {\rm H}(1s)
\end{equation}
\citep{Peart1985, Ferland1989}.
We assume that $n=3$ is statistically populated.

\NEW{The last two entries represent a few of the ``bands'' we report.
These are integrals of the emission, lines and continuum, over specified wavelength bounds.
The bands can be changed by the user by editing the file \texttt{continuum\_bands.ini}.
Currently the integrated emission is reported, 
which is equivalent to assuming a uniform instrumental sensitivity.
A number of bands, corresponding to a number of the more widely used
filter or spacecraft instrumental bands, are reported.
The two listed are ``TIR 1800m'', the integral from 500 \micron\ to 3100 \micron,
and ``TALL 10000A'', the integral from 1\e{-6} \micron\ to 1\e4 \micron.}

\subsection{XDRs and AGN}

Figure~\ref{fig:XdrAgnSed} shows that the XDR continuum is but a small part of the total SED
of an AGN.
The lighter line is the mean AGN SED derived by  \citet{Mathews1987} and built into \Cloudy.
The XDR SED is meant as a way to compute the conditions in the \hO\ region
that lies behind (as seen from the central object) the \hplus\ region where most
hydrogen-ionizing radiation is absorbed.
As \citet{Abel2005} stress, this may be a great oversimplification.

To check this, we computed two models of a ``Narrow Lined Region'' (NLR) cloud
near an AGN.  
These are lower-density dusty regions that contribute to the optical spectrum and
are likely to be ionized layers on the surface of larger molecular clouds (AGN3).
Many different sets of chemical abundances are built into \Cloudy.
We use a standard ISM gas-phase composition and a mixture of graphite and silicate grains
combined with an MRN size distribution. 
The clouds have a physical thickness corresponding to \NEW{a column density of
$N(\textrm{H}) =  3\times 10^{24} \pscm$, for} a point-source \Av\ of $\sim 10^3$ mag.
Galactic background cosmic rays were assumed.
We now adopt the \citet{Indriolo2007} mean \hO\ cosmic ray ionization rate 
of $2\times 10^{-16} \ps$ as the default Galactic background.

The AGN radiation field intensity was set with the dimensionless ionization parameter $U$,
defined as the ratio of hydrogen-ionizing photon to hydrogen densities (AGN3).
We adopt $\log U = -1.5$, a typical value deduced from the optical emission line spectrum
\citep{FergusonKoristaBaldwin1997}.
We normalized the XDR continuum to have the same 1 - 100 \keV\ flux,
83.18 \ergpscmps, as the AGN continuum.  
The hope is that an XDR computed with this SED and flux would be similar
to the \hO\ region in the AGN cloud.

The equation of state relates the gas density to other physical quantities such as the
kinetic temperature or radiation pressure (see Section \ref{sec:eos-momentum} above).
We assume constant total pressure.
This is very important for the AGN continuum which produces a hot ($\sim 1\e4 \K$) layer of ionized
gas on the surface of the cloud.
The illuminated face of the XDR cloud is predominantly atomic and warm ($\sim 1\e3 \K$)).
As a result, \NEW{for a given total hydrogen density} the gas pressures will differ by about 2 dex, 
the difference in temperature and particle density.
We want the densities in the \hO\ region to be comparable if we are to make a meaningful 
comparison between the two simulations. 
The pressure in the AGN simulation is 5.97\e{-8} \pressure.
We use the same pressure in the XDR. 
The hydrogen density and temperature at the illuminated faces of the AGN and XDR clouds is then
$10^4 \pcc$, $1.88\e4 \K$ and 
$2.33\e6 \pcc$, $2.29\e2 \K$ respectively.
Hydrogen in the AGN is fully ionized at this point while the XDR has 29\% \htwo.

\begin{figure}
\centering
\includegraphics[width=\linewidth]{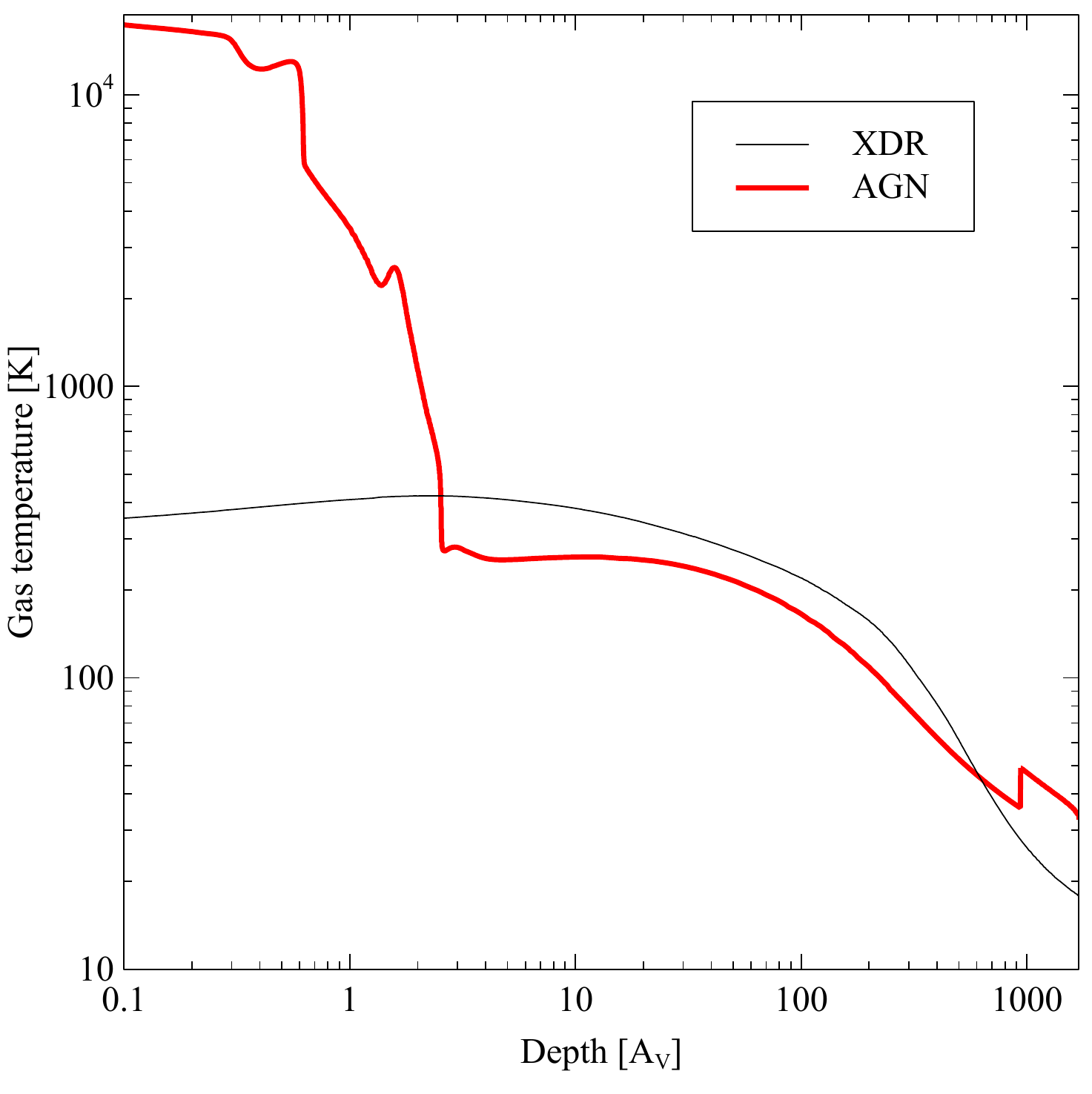}
\caption[Kinetic temperatures for XDR and AGN simulations]{
\label{fig:XdrAgnTemperature}
This compares the gas kinetic temperature for XDR and AGN clouds
with the same X-ray flux, the SEDs shown in Figure~\ref{fig:XdrAgnSed},
and the same total pressure.}
\end{figure}

\begin{figure}
\centering
\includegraphics[width=\linewidth]{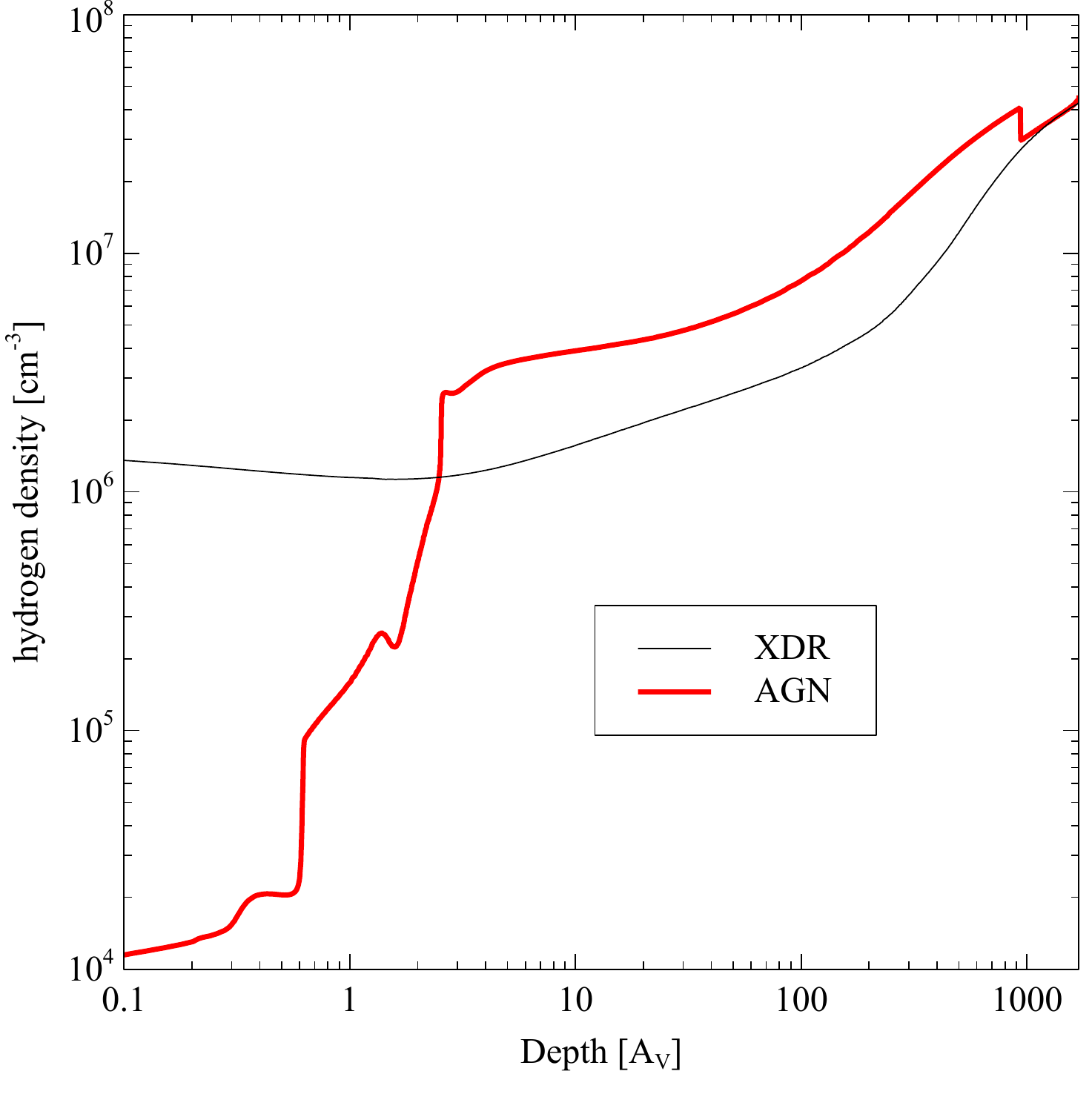}
\caption[The hydrogen density for XDR and AGN simulations]{
\label{fig:XdrAgnDensity}
The hydrogen density is shown for
the XDR and AGN clouds.}
\end{figure}

\begin{figure}
\centering
\includegraphics[width=\linewidth]{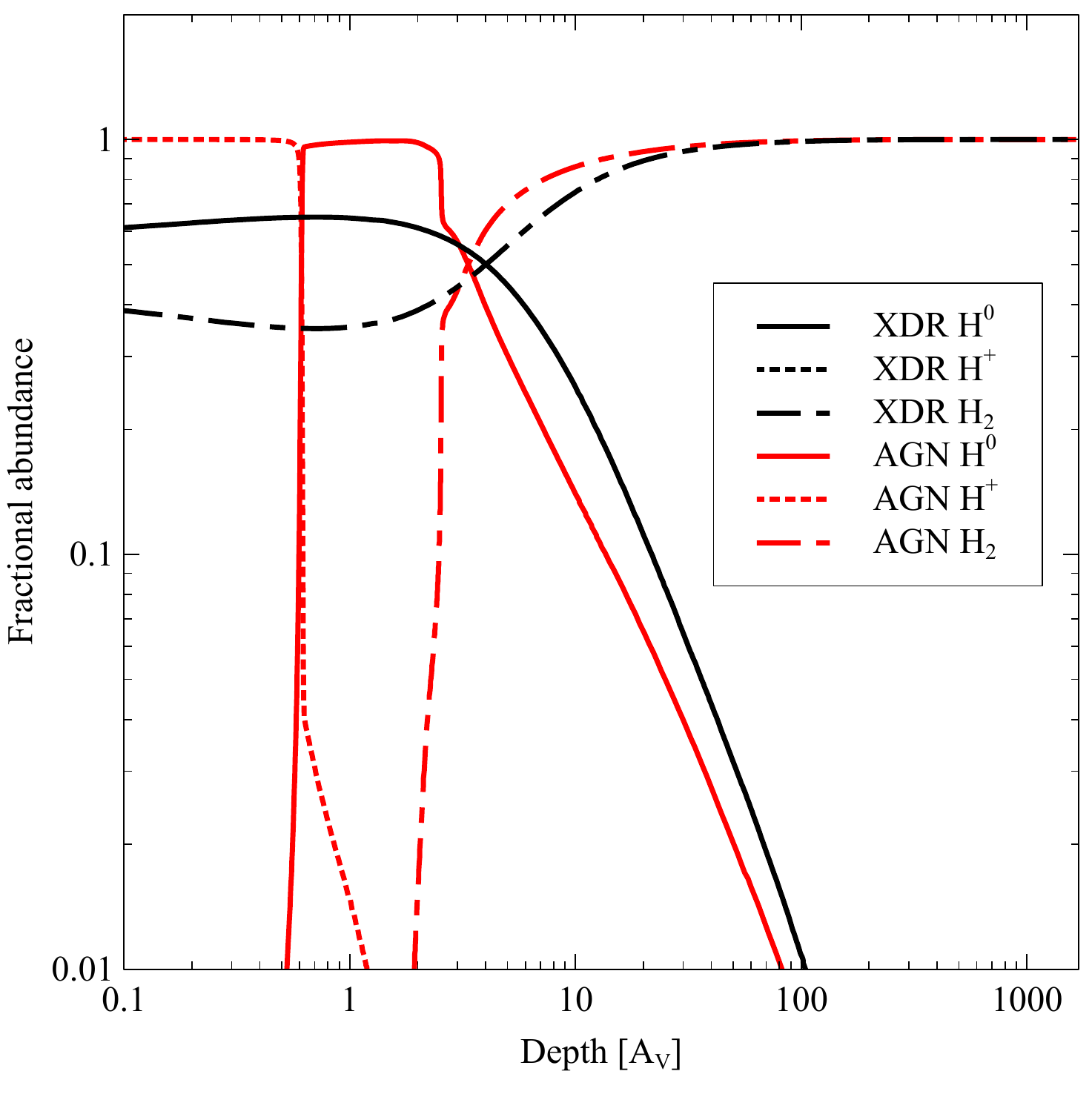}
\caption[Hydrogen fractions for XDR and AGN simulations]{
\label{fig:XdrAgnHstructure}
The hydrogen atomic, molecular, and ion fractions are shown for
the XDR and AGN clouds.}
\end{figure}

Figure~\ref{fig:XdrAgnTemperature} shows the gas kinetic temperature and Figure
\ref{fig:XdrAgnDensity} shows the hydrogen density as a function of depth from the 
illuminated face of the layer for the two scenarios.  
Depth is shown in terms of the point-source \Av\ to be consistent with other literature.  
Figure~\ref{fig:XdrAgnHstructure} shows the distribution of hydrogen in its various forms.

There is a very warm ionized layer in the AGN case due to the \hplus\ zone where 
hydrogen-ionizing photons are absorbed.  
This layer produces most of the emission from the cloud since the AGN SED peaks in the FUV and EUV.  
This emission, predominantly lines in the optical and UV, may be undetectable
if there is a large amount of surrounding dusty material.
The UV/optical emission would then be reprocessed by other clouds into IR emission.
In either case, the XDR continuum misses the majority of the continuum available for reprocessing.
We shall predict the emission emergent from the cloud we model, and do not consider
further reprocessing by other clouds in the system.

There are surprising large differences in the kinetic temperature in relatively shallow regions 
of the \hO\ region, which begins at about \Av\ $\sim 1$.  
The XDR produces a very flat temperature profile with $T \sim 300 - 400 \K$ being typical.  
The AGN produces a much warmer \hO\  layer at shallow depths with 
temperatures ranging from $T \sim 3000 \K$ to
$T \sim 200 \K$.
Soft X-rays that filter through the \hplus\ layer into the \hO\ regions produce the warm gas.  

\begin{figure}
\centering
\includegraphics[width=\linewidth]{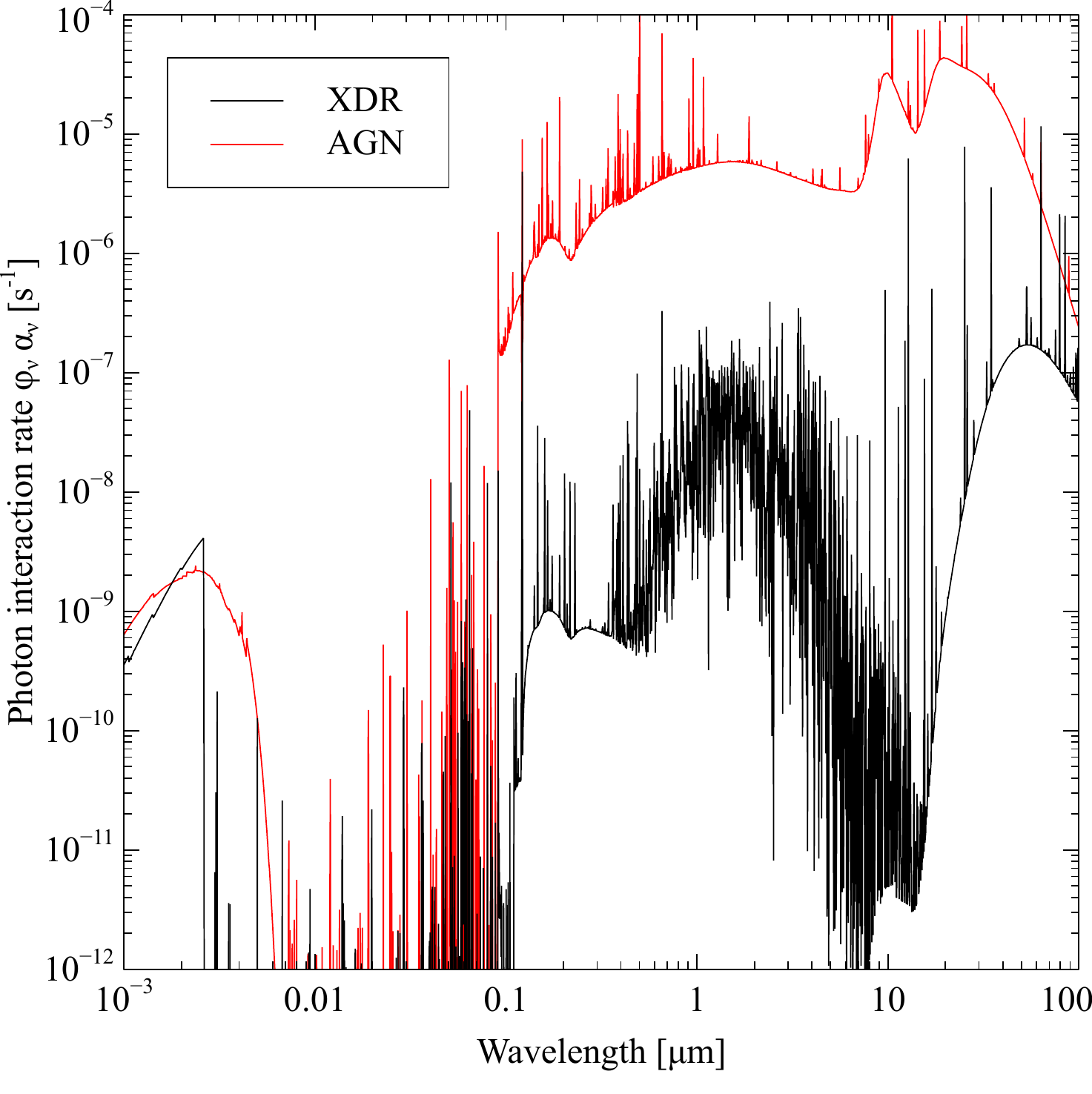}
\caption[The photon interaction rate at \Av$ = 1.5$ for XDR and AGN simulations]{
\label{fig:XdrAgnAv15}
The photon interaction rate at \Av$ = 1.5$ is shown for the XDR and AGN simulations.
This is the point where the AGN has a much warmer \hO\ region, 
produced by the emission from the adjacent \hplus\ region.}
\end{figure}

Figure~\ref{fig:XdrAgnAv15} shows the photon interaction rate at a depth of \Av $ = 1.5$,
the region where the temperature differences between \NEW{the \hO\ regions of} the XDR and AGN simulations are the largest.  
Roughly the same amount of energy is present in the region of the spectrum where the XDR 
incident continuum is defined, $1 - 100 \keV$,
although the transmitted AGN SED extends down to  lower energies.
This spectral region adds additional heating to the gas. 
The greatest differences are at $\lambda > 0.1 \micron$, where $\phi_{\nu} \alpha_{\nu}$
is about 3 dex larger in the AGN simulation.
This radiation heats the gas through grain electron photoejection, producing the much higher temperature.

Table~\ref{tab:XDRAGNlines} compares predicted intensities for the XDR and AGN simulations.
The XDR approximation does roughly agree with the AGN case for some MIR lines.
Standard XDR emission lines such as [\cii], [\oi], etc, are generally within factors of
$0.3 - 0.5 $ dex of one another.
Lines from higher ionization species, such as [\neiii] and [\siii] are bright in the AGN
but missing from the XDR due to the assumed SED.
The AGN produces far more total power since an AGN SED peaks at energies that are
not included in a standard XDR calculation.

\begin{table}[!t]
\centering
\caption{\NEW{Line intensities for XDR \& AGN simulations}}
\label{tab:XDRAGNlines}  
\tablecols{3}
\setlength\tabnotewidth{0.7\linewidth}
\newcommand\NODATA{\multicolumn{1}{c}{\nodata}}
\begin{tabular}{ c l l }
\toprule
Line& AGN\tabnotemark{a} & XDR\tabnotemark{a}\\ \midrule
\verb|H  1  6563A| & \(2.76       \) & \(4.24 \e{-3}\) \\
\verb|H  1  4861A| & \(8.61 \e{-1}\) & \(1.01 \e{-3}\) \\
\verb|H  1 1.875m| & \(3.44 \e{-1}\) & \(3.85 \e{-4}\) \\
\verb|O  2  3727A| & \(1.93 \e{-1}\) & \(1.34 \e{-7}\) \\
\verb|O  3  5007A| & \(9.99       \) & \NODATA         \\
\verb|N  2  6584A| & \(3.67 \e{-1}\) & \NODATA         \\
\verb|S  2  6720A| & \(5.62 \e{-1}\) & \(6.66 \e{-21}\) \\
\verb|H2   2.121m| & \(4.72 \e{-3}\) & \(4.16 \e{-3}\) \\
\verb|H2   17.03m| & \(2.20 \e{-2}\) & \(5.15 \e{-2}\) \\
\verb|H2   12.28m| & \(4.92 \e{-3}\) & \(2.40 \e{-2}\) \\
\verb|H2   9.662m| & \(2.68 \e{-3}\) & \(2.32 \e{-2}\) \\
\verb|C  1 609.2m| & \(1.03 \e{-4}\) & \(1.80 \e{-4}\) \\
\verb|C  1 369.7m| & \(5.61 \e{-4}\) & \(9.64 \e{-4}\) \\
\verb|C  2 157.6m| & \(9.48 \e{-3}\) & \(3.69 \e{-3}\) \\
\verb|NE 2 12.81m| & \(6.15 \e{-1}\) & \(2.51 \e{-1}\) \\
\verb|NE 3 15.55m| & \(1.35       \) & \(2.99 \e{-3}\) \\
\verb|SI 2 34.81m| & \(1.34 \e{-1}\) & \(1.82 \e{-1}\) \\
\verb|S  3 18.67m| & \(6.60 \e{-1}\) & \(1.69 \e{-7}\) \\
\verb|FE 2 1.644m| & \(5.15 \e{-3}\) & \(1.53 \e{-13}\) \\
\verb|C  1 609.2m| & \(5.48 \e{-5}\) & \(8.54 \e{-5}\) \\
\verb|C  1 369.7m| & \(3.83 \e{-4}\) & \(6.26 \e{-4}\) \\
\verb|C  2 157.6m| & \(1.41 \e{-2}\) & \(3.37 \e{-3}\) \\
\verb|O  1 63.17m| & \(5.65 \e{-1}\) & \(2.51 \e{-1}\) \\
\verb|O  1 145.5m| & \(2.71 \e{-2}\) & \(1.83 \e{-2}\) \\
\verb|Ne 2 12.81m| & \(8.83 \e{-1}\) & \(3.07 \e{-1}\) \\
\verb|Ne 3 15.55m| & \(2.46       \) & \(3.53 \e{-3}\) \\
\verb|Ne 3 36.01m| & \(1.66 \e{-1}\) & \(1.17 \e{-4}\) \\
\bottomrule
\tabnotetext{a}{\ergpscmps}
\end{tabular}
\end{table}

\begin{figure}
\centering
\includegraphics[width=\linewidth]{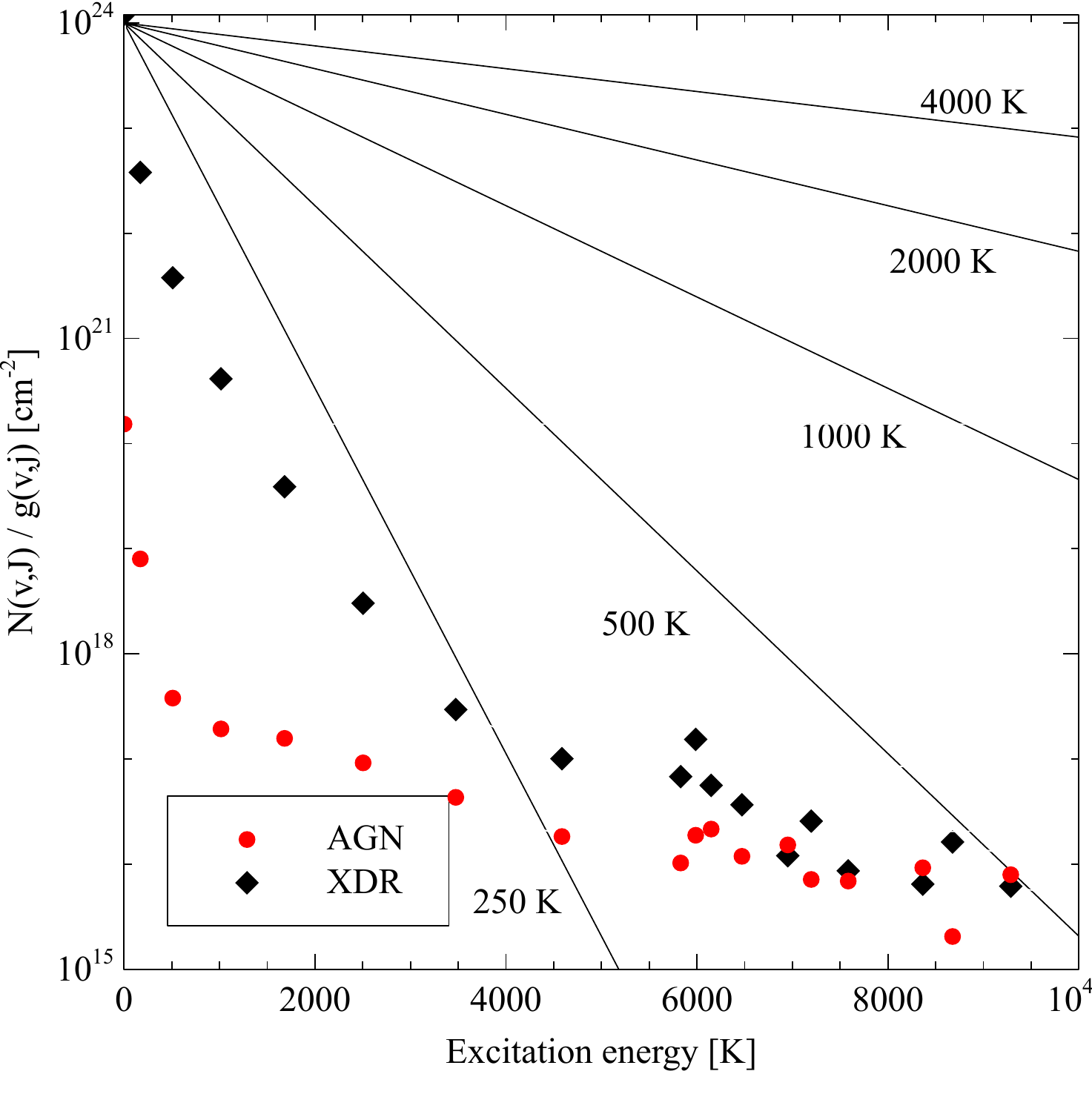}
\caption[\htwo\ excitation diagram]{
\label{fig:H2Excitation}
\htwo\ excitation diagram.  The x-axis is the excitation energy of the v,J levels expressed in \K.
The y-axis is the predicted column density of the v,J level divided by its statistical weight.
The lines indicate thermal distributions at various temperatures.}
\end{figure}

\htwo\ excitation diagrams are often used to probe physical conditions in warm molecular regions.
Figure~\ref{fig:H2Excitation} shows this diagram for the two models.
The overall distribution of \NEW{higher populations, with $T_{exc} > 4000 \K$, are similar.
Lower populations indicate cooler gas for the XDR, as suggested by Figure~\ref{fig:XdrAgnTemperature}.}
As is typical for such diagrams, lower levels, which can be excited by cooler gas,
indicate lower temperatures than the high levels,
which are only excited in warmer regions \NEW{or by continuum pumping}.
The populations below 2000 \K\ \NEW{have a $\sim 250 \K$ Boltzmann distribution for the XDR}, a temperature
something like that of the \htwo\ region in Figure~\ref{fig:XdrAgnHstructure}.

\begin{figure}
\centering
\includegraphics[width=\linewidth]{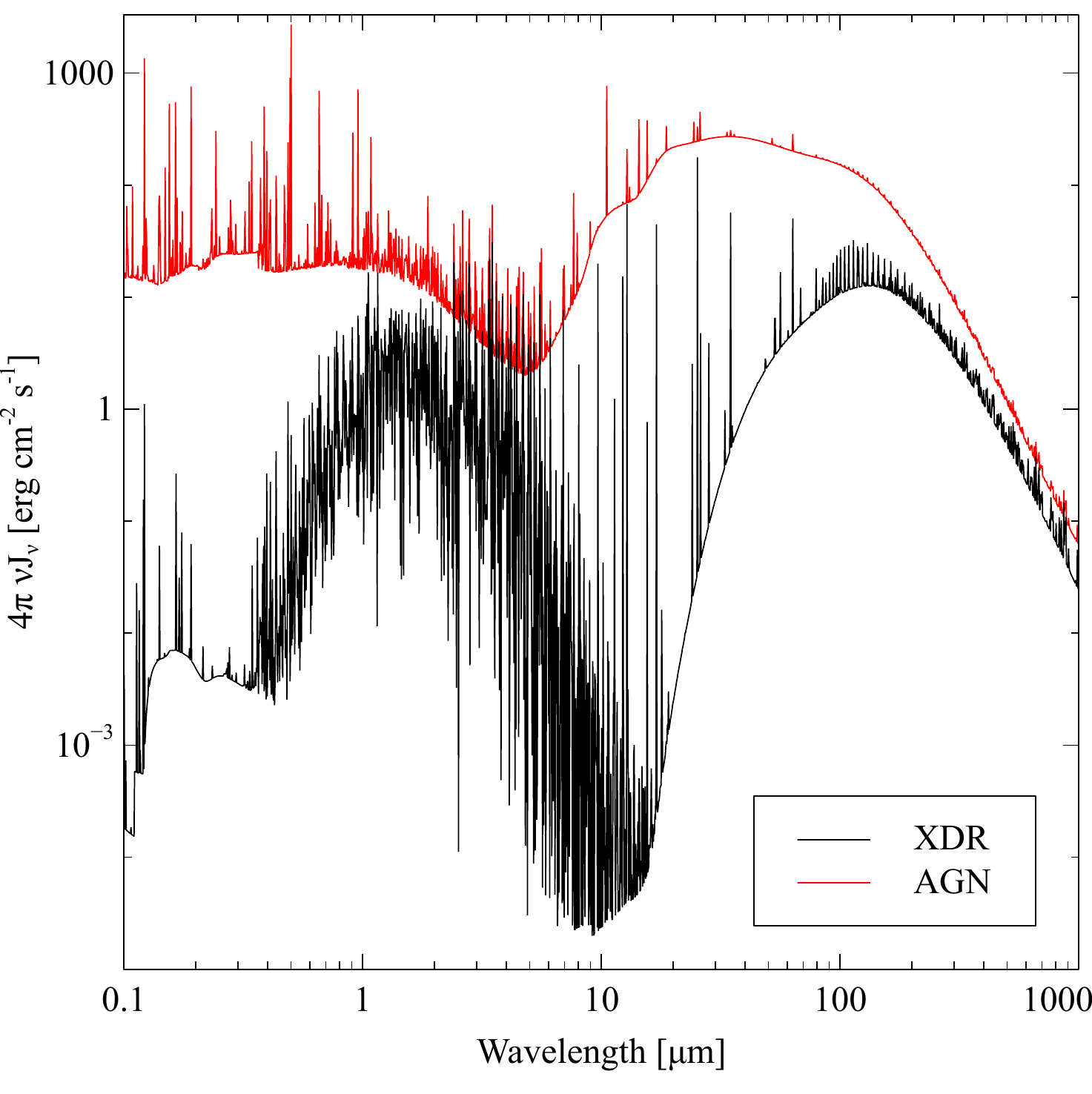}
\caption[The emergent spectrum for the XDR and AGN]{
\label{fig:XdrAgnEmergentCont}
\NEW{The spectra emitted by the XDR and AGN simulations are shown.}}
\end{figure}

\NEW{Figure~\ref{fig:XdrAgnEmergentCont} compares the XDR and AGN emergent spectra.
The XDR is a strong source of molecular emission, filling wavelengths longward
of $\sim 100 \micron$.  
The large ``bump'' of emission between $\sim 1 - 10 \micron$ is largely produced by \htwo\ lines.
Molecules are prominent since little UV light is present to dissociate them,
as shown in Figure~\ref{fig:XdrAgnAv15}.
The AGN produces strong optical emission due to the warm \hplus\ layer,
and atomic and ionic emission in the IR.
The dust emission is considerably warmer in the AGN case due to heating by
the UV and optical radiation field.}

\subsection{Physical conditions over an extreme range of matter and photon density}
\label{sec:GridExtreme}

The two previous sections highlight the types of physics that has been a particular
emphasis in the code's development since F98, dusty atomic and molecular regions.  
\Cloudy\ is designed to faithfully simulate physical processes that occur in 
the full range of density and temperature
encountered in interstellar clouds, accretion disks, or dense accretion flows.
To demonstrate this range we computed (using the \textit{grid} command described above)
the properties of a unit cell of
a photoionized cloud over a very wide range of density and intensity of the incident radiation field. 
Such tests are important because they show that  predictions agree with analytical theory
for asymptotic limiting cases. 

\begin{figure*}[t!]
\centering
\includegraphics[width=0.8\linewidth]{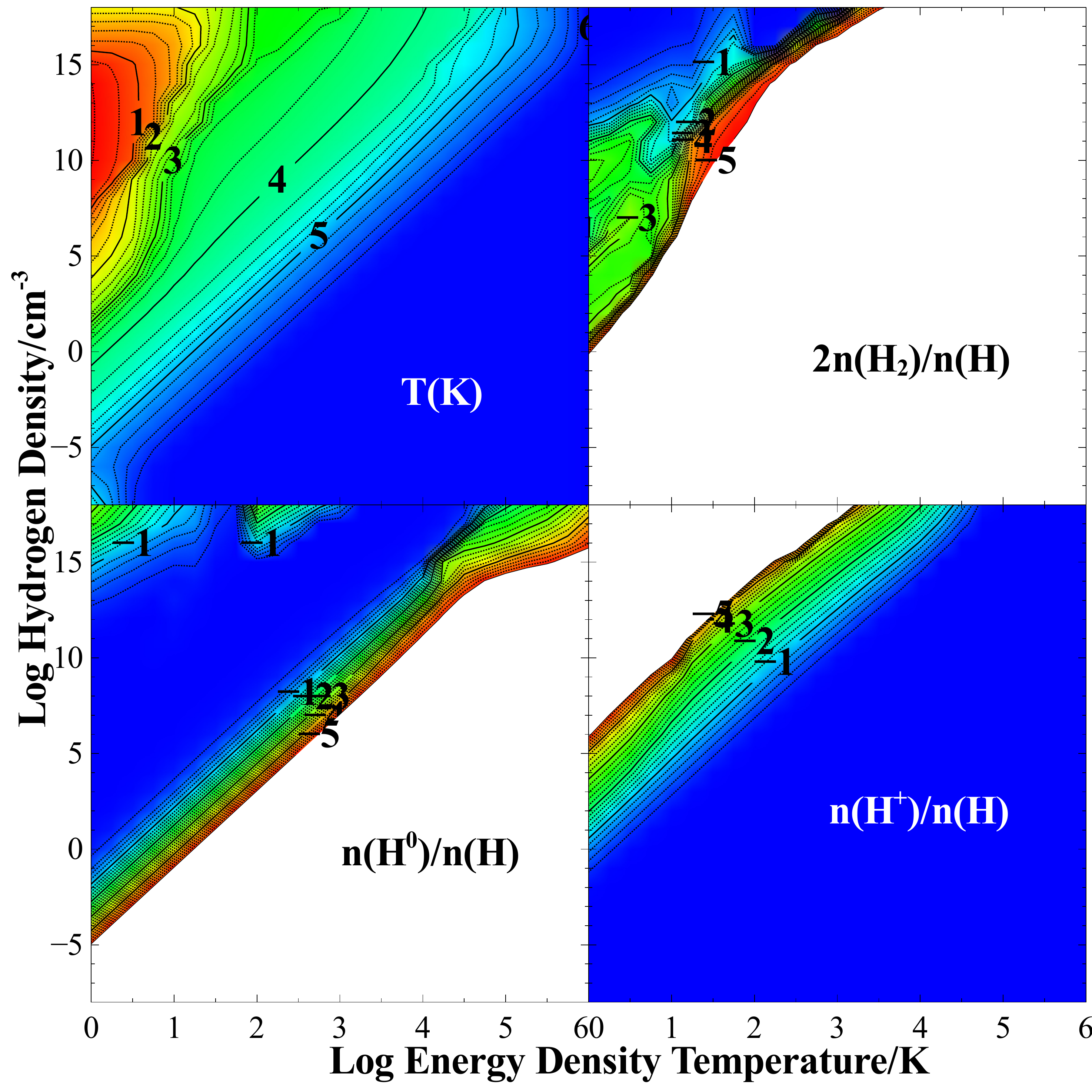}
\caption[Gas-kinetic temperature across a wide range of  
density and intensity]{
\label{fig:GridExtreme}
The physical properties of an irradiated cell of gas are is shown across a wide range of  
gas density and radiation field intensity.   
The upper left panel shows the log of the kinetic temperature as a function of gas
density (the vertical axis) and the energy density of the radiation field
(the horizontal axis).
The other three panels show logs of the hydrogen molecular fraction, $2n(\htwo)/n({\rm H})$,
and atomic and ion fraction.} 
\end{figure*}

The gas has solar composition (without grains)
and is illuminated by a blackbody with a $T_{BB} = 10^6$ K color temperature but with a variable intensity.  
The hydrogen density ranges from $10^{-8}\pcc$, below that of the IGM, to $10^{18} \pcc$, 
a density that is typical of the atmospheres of some stars or accretion disks.  
The horizontal axis is the intensity of the black body given as the energy-density temperature, 
$T_u = (u/a)^{1/4}$ K, where \(u\) is the total energy density in all wavelengths [erg cm$^{-3}$] 
and \(a\) is the Stefan radiation-density constant.
This range in $T_u$ includes environments extending from the 
Intergalactic Medium (IGM) to deep layers within a star.  
Most clouds encountered in astrophysics have a gas and energy density that
lies somewhere in  Figure~\ref{fig:GridExtreme}.

The upper left panel of Figure~\ref{fig:GridExtreme} shows the predicted gas kinetic temperature.
This ranges from low values typical of cold molecular gas 
(the upper left-hand corner of the figure) to high values in the highly ionized right end.  
The gas temperature closely approaches $T_{BB}$ as $T_u \rightarrow T_{BB}$,
as it must from thermodynamics.
The right edge of the Figure corresponds to
a radiation field in strict thermodynamic equilibrium (STE) since $T_u = T_{color}$.  
There is no lower bound to the gas kinetic temperature but we do see that generally
$T_{kin} > T_u$ since the gas is not a perfect radiator.
The lowest temperatures occur for the denser gas at the lowest $T_u$.

The remaining panels of Figure~\ref{fig:GridExtreme} show the state of hydrogen.
Only fractions $n(X)/n({\rm H})> 10^{-5}$ are plotted for simplicity.
Bands of constant ionization parameter $U$ are the diagonals running from lower left to upper right.
The gas is highly ionized in the lower right half high-$U$ region.
Moving to lower $U$, going from the lower right corner towards the upper left,
the gas becomes first atomic then molecular.
In low-$U$ molecular regions the chemistry occurs totally in the gas phase since grains are not present.

\begin{figure}
\centering
\includegraphics[width=\linewidth]{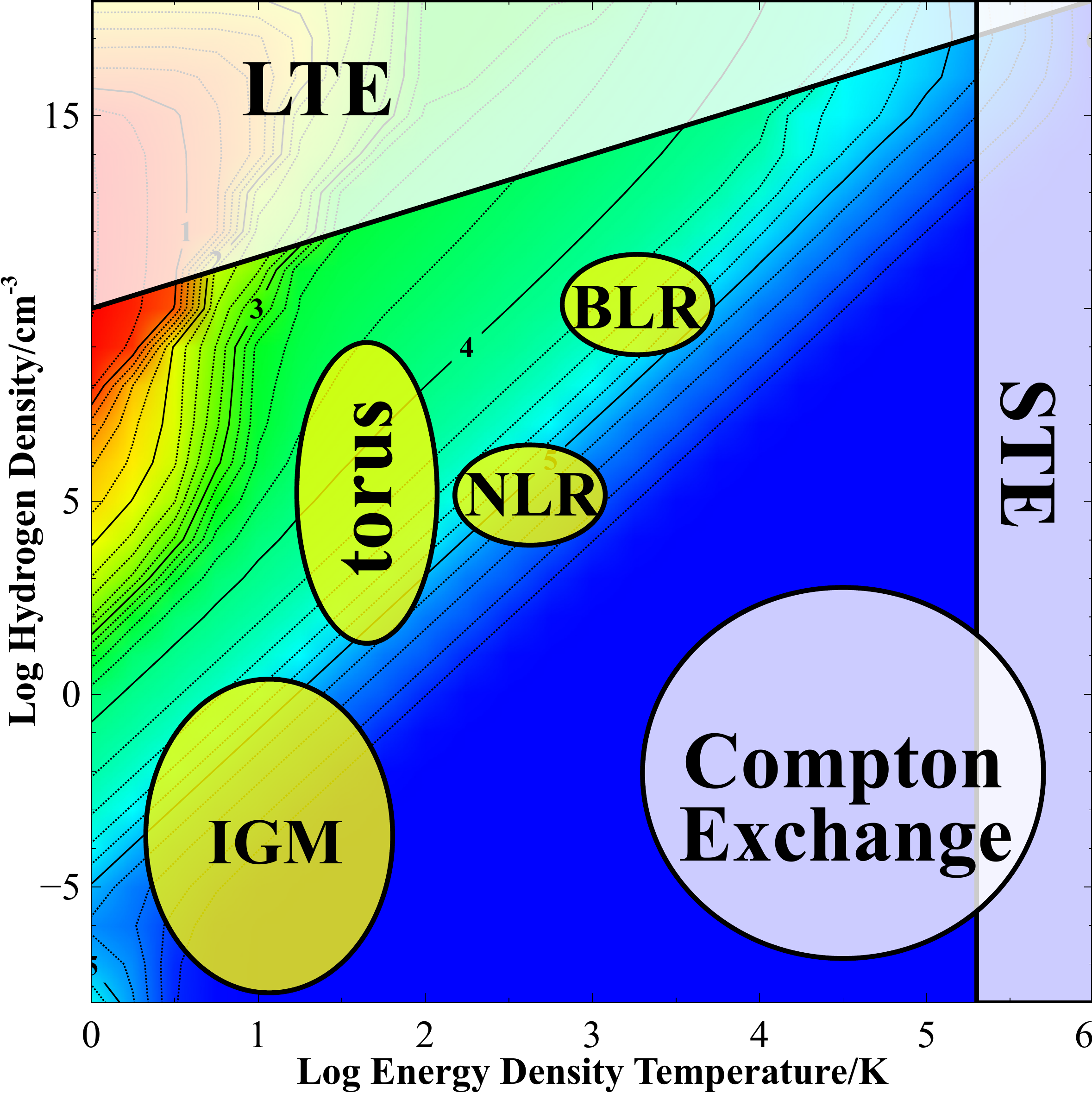}
\caption[Gas-kinetic temperature across a wide range of  
density and intensity]{
\label{fig:GridExtremeCartoon}
This panel identifies some physical and thermodynamic limits \NEW{(in white)} 
and shows where some regions in Active Galactic Nuclei are located \NEW{(in yellow)}, 
for the calculations shown in Figure~\ref{fig:GridExtreme}.
A wide range of densities, and various energy-density temperatures of the $10^6 \K$
blackbody, are shown.}
\end{figure}

Figure~\ref{fig:GridExtremeCartoon} is an annotated version of  Figure~\ref{fig:GridExtreme} 
summarizing some physical limits
and showing locations of some astronomical objects.
This Figure is meant to illustrate the physics occurring in various combinations of
density and radiation field, and is not meant to be rigorous.

Gas in the high-density region of the Figure will be in local thermodynamic equilibrium, LTE,
\citep{Mihalas1978, Rutten2003} when
the density is high enough for thermal collisions to control  the ionization and level populations.
Levels are said to be in LTE if their level populations are given by Boltzmann statistics
for the local gas kinetic temperature.  
The radiation field may, or may not, be a black body at this temperature.
Notice that only some levels of an ion may be in LTE.  In general higher \NEW{electronic} levels
come into LTE at low densities, because of larger collision cross sections and lower
transition probabilities.
In C13 only ions of the H- and He-like iso-sequences have enough high levels to go to LTE.
Higher densities are needed to go to LTE at larger $T_u$ for two reasons.
Both the level of ionization and the illuminating radiation field increase with increasing $T_u$.  
Higher densities are needed if collisions are to dominate rates for level populations.

The gas is said to be in strict thermodynamic equilibrium (STE) when 
the ionization, level populations, and radiation field are given by the same temperature.
This occurs at the right edge of Figure~\ref{fig:GridExtremeCartoon},
where $T_u \rightarrow T_{BB}$.
Our test suite includes many cases confirming that predictions go to the
LTE and STE limits where expected.

The temperature in the lower-right quadrant is determined by Compton energy exchange 
\citep{Ferland1988}, which drives $T_{kin} \rightarrow T_{BB}$.  
Here photon - electron collisions dominate the energy exchange and the gas temperature
approaches a value determined by the SED of the radiation field.
Compton exchange dominates when there is little absorption opacity,
which is true for the highest-$U$ regions in the lower right of the diagram.

Classical Str\"{o}mgren photoionization (AGN3) operates in the mid-$T_u$, low to mid density,
regions of the Figure. 
Here the approximation that most atoms are in the ground state and that all
recombinations eventually reach ground (the equivalent two-level atom) is valid.

Grains tend to equilibrate at temperatures near $T_u$ and have sublimation temperatures
around $1 - 1.5\e3 \K$.
They can only exist in the left third of the diagram.

Figure~\ref{fig:GridExtremeCartoon}
shows where some of the emission-line regions of Active Galactic Nuclei (AGN) are located.
The narrow-lined region (NLR) may be
molecular clouds irradiated by the radiation field of the AGN. 
The BLR, 
likely the skin of an accretion disk near the
supermassive black hole, lies between the LTE and Str\"{o}mgren regimes.  
This environment is dense enough for there to be significant populations of
excited states.  Photoionization and collisional ionization from these states, 
and the radiative transfer effects produced by their large populations,
all make this a computationally challenging environment.
The molecular torus, the dusty warm molecular gas that exists outside the
accretion disk and creates the AGN 1 / AGN 2 distinction (AGN3), 
lies along the left.
The intergalactic medium (IGM) has lower gas density and is illuminated by a weak radiation field.

Figures \ref{fig:GridExtreme} and \ref{fig:GridExtremeCartoon} show that such diverse phenomena 
as the IGM, AGN molecular torus, the NLR,
and the BLR are simply manifestations of different regimes of atomic and molecular physics.  
This is the approach we take.  
If the microphysics is done at an elementary level the macrophysics will follow.

\section{A look forward}

The development of \Cloudy\ continues.
The goal is a true simulation of the microphysics and spectrum of
gas and dust over the range of conditions shown in Figure~\ref{fig:GridExtremeCartoon}.
Our calculations have always been limited by processor power, and many important
physical processes are at the forefront of research in atomic, molecular, or grain physics. 
New simulations, which will offer better insight into what happens in front of our telescopes,
will be possible with faster computers, improved numerical methods, and better physical data.

The code infrastructure is being improved.
We had developed our own database for physical processes,
on an ad hoc basis, and embedded it within the C++ source code.
This makes it very difficult to update the database as improvements or
extensions occur.
We are now well into moving physical data into external databases which are
parsed when the code is initialized.  
This effort should be complete by the next release.
This external database will be public, along with the rest of \Cloudy.

As described in Section \ref{parallel}, two commands make it possible
to perform a large number of simulations in parallel using MPI.
This type of ``embarrassingly parallel'' calculation is ideal for distributed memory systems.

Shared memory systems should be easier to program and might be used to make single
models faster.
As described throughout this paper, a calculation simultaneously and self-consistently
solves a large number of relatively modest problems.  There is no ``long pole in the tent''
to go after in searching for tasks to make parallel.  There are many cross-dependencies
between the physical parameters we calculate. Just one example: to calculate the ionization
structure you need to know the radiation field, but to calculate the radiation field you
need to know the ionization structure. There are many more dependencies like this one, forcing
us to use iterative schemes in many places. These make parallellizing the code a lot harder.
Worse, the data layout in memory is often less than optimal,
resulting in poor cache utilization.  Some changes have been made to
improve cache locality and the potential for vectorization, but more work remains to be done.
The speed of calculations on modern CPU architectures is often limited by memory bandwidth rather than compute speed.
We are still considering how to take better advantage of today's multi-core processors.

To put all this in perspective, our \textit{pn\_paris} test case, one of the simulations 
from the 1985 Paris meeting, took about a minute to compute on large
mainframes at the time.
Today the simulation still requires about a minute, despite the astonishing increase
in computer power in the past 28 years.
Today's simulation includes many more physical processes, far better emission models,
and is a much more robust model of the real nebula.

The grain physics will be improved,  driven by the remarkable advances
from recent infrared space missions.
We will include
more grain surface reactions, thought to be important in forming complex molecules.
Grain opacities, especially for PAHs,  depend on
charge and temperature and are not a constant for a particular material and size.
Finally, 
radio emission from spinning grains can be important and is being developed.

Filaments in cool-core clusters of galaxies are thought to be excited by penetrating
energetic particles from the surrounding hot intracluster medium
\citep{FerlandFabianEtAl2009, Fabian.A11The-energy-source-of-the-filaments-around}.
Our treatment of cosmic ray or energetic particles does not now include attenuation
\citep{FerlandMushotzky1984},
which will depend on an uncertain magnetic field geometry.
Theories for cosmic ray transport do exist 
\citep{Padovani.M09Cosmic-ray-ionization-of-molecular-clouds} and may be incorporated.

Tests shown in previous papers, and demonstrated in our test suite,
show that species treated with our iso-sequence model go to LTE in the
high radiation or particle density limits.
These include all one and two-electron species.
Other ions are treated assuming equivalent two-level systems,
as described above, and cannot to go LTE.
This is the greatest weakness in our simulations at high densities.
We intend to extend the iso-sequence approach to more species, using
accurate atomic databases to model lower levels.
We can now use both Chianti and Stout, our new external database.
Chianti does not include subordinate collisions and so cannot go to LTE.
It was intended for relatively low densities.
Our Stout database includes all collisions.  Neither extends to high
enough energy levels for collisional coupling to the continuum and LTE to occur.
These models will have to be supplemented with higher Rydberg levels to allow 
the appropriate high-density behavior to occur.

\NEW{Much work remains to be done.  A true simulation of the
physical state of matter over the extremes of conditions found in astrophysics
is the first step in understanding the messages in the light we observe.
This goal is within sight.}

\section*{DEDICATION}

\textit{We dedicate this paper to Manuel Peimbert.  Through his leadership the
traditions established by Menzel, Baker, \NEW{Aller}, Str\"{o}mgren, Osterbrock \& Seaton
are being carried on, in Mexico.}

\acknowledgments

GJF acknowledges support by NSF (1108928; and 1109061), NASA (10-ATP10-0053, 
10-ADAP10-0073, and NNX12AH73G), and STScI (HST-AR-12125.01, 
GO-12560, and HST-GO-12309). PvH acknowledges support from the Belgian Science
Policy Office through the ESA Prodex program.  WJH acknowledges funding from DGAPA-UNAM, Mexico, through grant PAPIIT IN102012.

\bibliography{LocalBibliography,./common/bibliography2}		

\end{document}